\documentclass[twocolumn]{aastex631}
\usepackage{booktabs}
\usepackage{amsmath}

\newcommand{\mtrgb}{$m_{\rm TRGB}$}
\newcommand{\Mtrgb}{$M_{\rm TRGB}$}

\newcommand{\gaia}{\textit{Gaia}}

\newcommand{\ogle}{\textit{OGLE}}

\newcommand{\mic}{$m_{I,\mathrm{syn}}$}
\newcommand{\mih}{$m_{\mathrm{F814W,syn}}$}

\newcommand{\mio}{$m_{I,\mathrm{OGLE}}$}

\newcommand{\Mih}{$M_{\mathrm{F814W,syn}}$}

\newcommand{\Mio}{$M_{I,\mathrm{OGLE}}$}

\newcommand{\Ml}{$M_{\lambda}$}

\newcommand{\Mi}{$M_I$}

\newcommand{\sigs}{$\sigma_s$}
\newcommand{\sigmc}{$\sigma_{\mathrm{MC}}$}
\newcommand{\sigphot}{$\sigma_{\mathrm{phot}}$}
\newcommand{\sigt}{$\sigma_{\mathrm{TRGB}}$}
\newcommand{\Mt}{$M_{\mathrm{TRGB}}$}
\newcommand{\mt}{$m_{\mathrm{TRGB}}$}

\newcommand{\mudiffogle}{$\Delta \mu_{\mathrm{SMC-LMC,}\mathrm{OGLE}}$}
\newcommand{\mudiff}{$\Delta \mu_{\mathrm{SMC-LMC}}$}
\newcommand{\varstars}{\texttt{SARGs}}
\newcommand{\sargs}{\texttt{SARGs}}
\newcommand{\AllStars}{\texttt{Allstars}}
\newcommand{\Ad}{\texttt{Aseq}$^\dagger$}
\newcommand{\Aseq}{\texttt{A-sequence}}
\newcommand{\Bseq}{\texttt{B-sequence}}

\begin{document}

\title{Calibrating and standardizing the Tip of the Red Giant Branch in the Small Magellanic Cloud\\ using small-amplitude red giants}
\shorttitle{Calibrating and standardizing the SMC's TRGB using SARGs}
\shortauthors{N.W.~Koblischke \& R.I.~Anderson}

\author[0000-0001-5396-5824]{Nolan W. Koblischke}
\affiliation{Institute of Physics, \'Ecole Polytechnique F\'ed\'erale de Lausanne (EPFL), Observatoire de Sauverny, 1290 Versoix, Switzerland}
\affiliation{David A. Dunlap Department of Astronomy and Astrophysics, University of Toronto, 50 St. George Street, Toronto, Ontario, M5S 3H4, Canada}
\email{nolan.koblischke@mail.utoronto.ca}

\author[0000-0001-8089-4419]{Richard I. Anderson}
\affiliation{Institute of Physics, \'Ecole Polytechnique F\'ed\'erale de Lausanne (EPFL), Observatoire de Sauverny, 1290 Versoix, Switzerland}
\email{richard.anderson@epfl.ch}
\correspondingauthor{Richard I. Anderson}

\begin{abstract}
    We investigate the absolute calibration of the Tip of the Red Giant Branch (TRGB) in the Small Magellanic Cloud (SMC) using small amplitude red giant stars (SARGs) classified by the Optical Gravitational Lensing Experiment (OGLE). We show that all stars near the SMC's TRGB are SARGs. Distinguishing older and younger RGs near the Tip according to two period-luminosity sequences labeled A and B, we show many similarities among SARG populations of the LMC and the SMC, along with notable differences. Specifically, SMC SARGs have shorter periods due to lower metallicity and smaller amplitudes due to younger ages than LMC SARGs. We discover two period-color relations near the TRGB that span all \Aseq\ and \Bseq\ stars in the OGLE-III footprints of the SMC and LMC, and we investigate using periods instead of color for TRGB standardization. Using variability derived information only, we trace the SMC's age and metallicity gradients and show the core to be populated by younger, more metal rich RGs.  
    The \Bseq\ yields the brightest and most accurate  calibration (\Mih$= -4.057 \pm 0.019 (\mathrm{stat.}) \pm 0.029 (\mathrm{syst.})$\,mag), which we use to measure the distance modulus difference between the Clouds and investigate metallicity effects. Distance measurements not informed by variability should employ the \sargs-based calibration based on all stars near the tip (\Mih$= -4.024 \pm 0.041 (\mathrm{stat.}) \pm 0.029 (\mathrm{syst.})$\,mag). Our work highlights the impact of RG population diversity on TRGB distance measurements. Further study is needed unravel these effects and improve TRGB standardization.
\end{abstract}


\section{Introduction \label{sec:intro}}

The Tip of the Red Giant Branch (TRGB) is an empirical feature in the color-magnitude diagrams of old stellar populations that serves as an important standard candle for determining luminosity distances \citep{Lee1993}. Indeed, the TRGB is the most commonly applied stellar standard candle for measuring extragalactic distances \citep[e.g.,][]{Anand2021} thanks to the high prevalence of evolved metal-poor stars in most galaxies out to a few tens of Mpc. This renders the TRGB a useful tool for measuring the Hubble constant via an extragalactic distance ladder, either as calibrators of type-Ia supernovae \citep[e.g.,][]{Freedman2021,Anand2022,Scolnic2023} or of surface brightness fluctuations \citep{Anand2024sbf}, cf. also the recent review by \citet{LiH0book2024}. Astrophysically, the TRGB feature is understood to be caused by the helium flash that abruptly inverts the evolutionary direction of first-ascent low-mass red giant branch (RGB) stars  \citep[$M < 2.2 M_{\odot}$]{Sweigart1978,Salaris2002}. The rapidity of the He flash creates a near discontinuity in the luminosity function of red giant stars, which is in practice frequently contaminated by younger and higher-luminosity asymptotic giant branch (AGB) stars.

As with any standard candle, both calibration and standardization are required in order to measure accurate distances using the TRGB \citep{Anderson2024book}. Standardization involves algorithmic subtleties \citep{Madore2009,Makarov2006,Hatt17,Wu2022} as well as corrections for astrophysical differences. For example, differences in chemical composition affect both the luminosity and the shape of the TRGB feature and vary according to the photometric bands considered. In particular, the $I-$band is known for its particularly flat TRGB that provides best consistency for distance estimates. In $I-$band, several studies have reported that higher metallicity leads to dimmer TRGB magnitudes \citep{Freedman2020, Rizzi2007, Hoyt2023}. Age differences are not usually considered due to lack of reliable information, and an expectation that age differences for very old red giants (several Gyr and older) should be small \citep[e.g.][]{Salaris2005,Serenelli2017}. The calibration of the TRGB absolute magnitude requires knowledge of geometric distances to stars undergoing the core Helium flash. The geometric distances currently available to this end are trigonometric parallaxes from the ESA \gaia\ mission \citep{GaiaMission,GaiaEDR3plx}, the distances of the Magellanic Clouds determined using evolved detached eclipsing binaries \citep{Pietrzynski19,Graczyk2020}, and the $H_{2}O$ megamaser distance to NGC\,4258 \citep{Reid2019}. Among these possibilities, the Magellanic Clouds benefit from a particularly comprehensive and long-term observational dataset that allows detailed studies of red giant (RG) populations. In particular, the Optical Gravitational Lensing Experiment (OGLE) has provided more than decade-long time series observations that allow to map even small-amplitude variability for millions of stars in the Magellanic Clouds \citep{Udalski08,Soszynski2004,Soszynski2011}.

Using photometry, reddening maps, and variability information delivered by OGLE, \citet[henceforth: A24]{Anderson2024} recently showed that virtually all stars near the TRGB in the LMC are small amplitude red giant stars (\sargs) that exhibit multi-modal long-period variability at the level of a few hundredths of a magnitude. Intriguingly, the dominant pulsation mode of \sargs\ allows to distinguish between younger and older red giant sub-populations in the Large Magellanic Cloud (LMC), which yield significantly different values for the TRGB magnitude. Hence, variability provides urgently needed information to distinguish ages and to probe systematics of the TRGB according to variability-selected RG sub-populations. While the amplitudes of \sargs\ are arguably too small to be readily measured beyond the local group, it is nonetheless likely that \sargs\ observed in nearby galaxies will allow to develop standardization methods for improving the accuracy of TRGB distances. 

In this \textit{article}, we build on A24 and investigate the TRGB calibration based \sargs\ in the Small Magellanic Cloud (SMC). We seek to further understand how variability can inform TRGB standardization in order to achieve the best possible absolute calibration for extragalactic distance measurements. The structure of the \textit{article} is as follows. Section\,\ref{sec:dataandmethods} describes the input data used, sample selection, and recalls the method developed in A24. An overview of statistical and systematic uncertainties is provided in Section \ref{sec:sys_unc}. Section\,\ref{sec:results} first presents apparent TRGB magnitudes determined using SMC samples and compares differences among variability-selected subsamples  (Sect.\,\ref{sec:mtrgb}).
In turn, Sect.\,\ref {sec:spatial} considers spatial variations and derives period-color relations for \sargs\ on two long-period variable sequences near the RGB tip (Sect.\,\ref{sec:periodcolor}). \sargs\ are used to measure the difference in distance modulus between the SMC and the LMC, $\Delta \mu_{\mathrm{SMC-LMC}}$, in Sect.\,\ref{sec:mudiff}, and the absolute magnitude of the TRGB is calibrated in Sect.\,\ref{sec:absmag}, which further considers metallicity differences traced by differences in pulsation periods.  Additional discussion of these results and of the variable TRGB is presented in Section \ref{sec:discussion}. The final Sect.\,\ref{sec:conclusions} summarizes our results and presents our conclusions.

\section{Data and Methods}\label{sec:dataandmethods}

\subsection{Sample Definitions\label{sec:samples}}

\begin{figure*}[ht!]
    \includegraphics[width=1\textwidth]{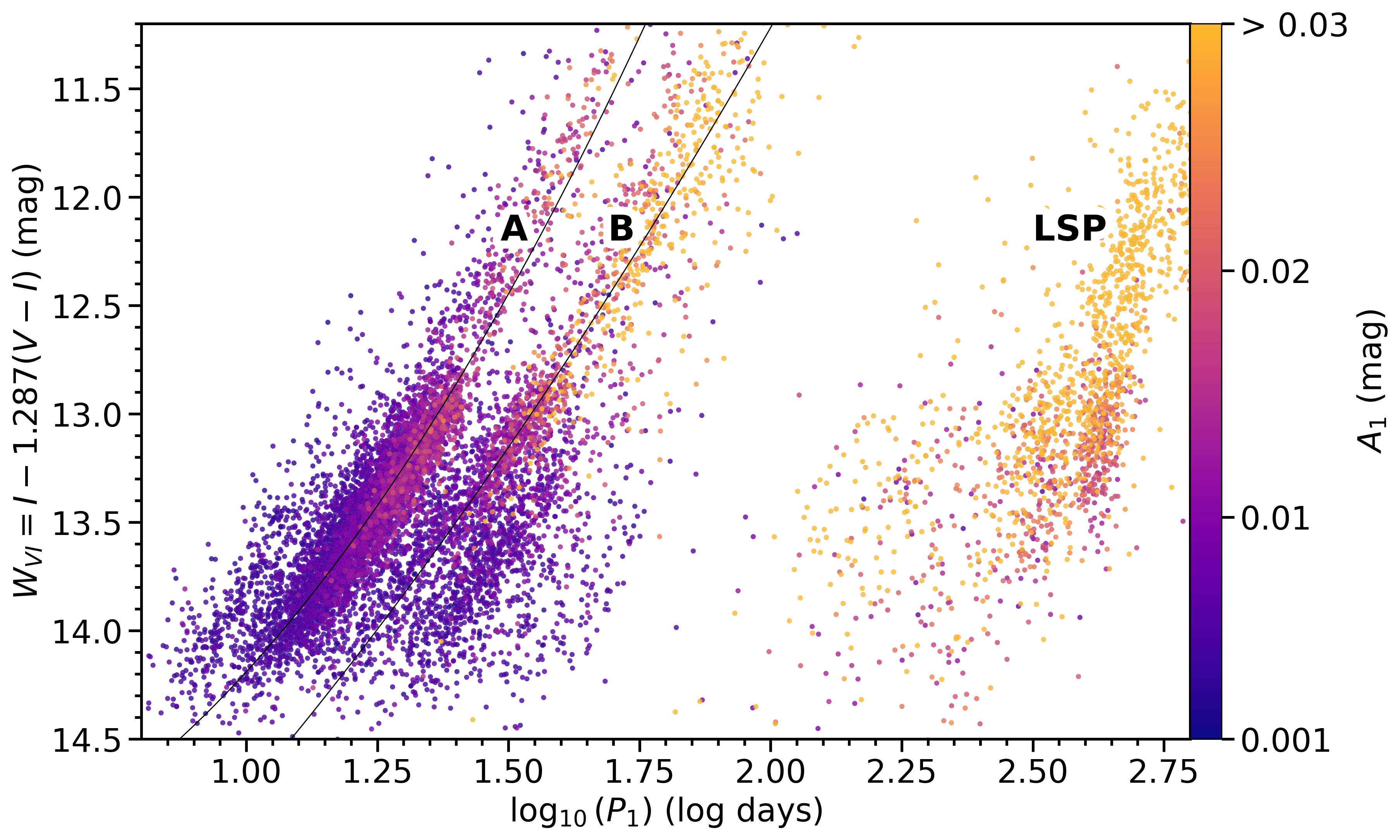}
    \caption{The period-luminosity relations of small amplitude red giants in the SMC in reddening-free Wesenheit magnitudes computed assuming $R_V = 2.7$ and using the primary pulsation period $P_1$. Black lines indicate the 2nd order polynomial fits used to select a $3\sigma$ band around the sequences. Points are colored by the amplitude of $P_1$. Sequences A and B are labeled along with Long Secondary Period (LSP) \citep{Nicholls2009}. The origin of the LSP phenomenon is not well understood, with theories ranging from binarity, dust, and non-radial pulsation \citep[e.g.,][]{2021ApJ...911L..22S,Pawlak2024}. Curiously, stars on sequence B are more likely to have a smaller-amplitude mode on the LSP sequence.}
    \label{fig:PL_AllSequences}
\end{figure*}

We considered four samples made up of RG stars in the OGLE-III footprint of the SMC as well as the LMC samples described in A24. The sample of all RG stars is henceforth referred to as \AllStars. The \sargs\ sample is a subset of \AllStars, whose variability was reported and classified in the OGLE-III catalog of long-period variables \citep{Soszynski2011}. The SMC sample of \sargs\ is significantly smaller than the LMC: we started with 16,810 \sargs\ in the SMC \citep{Soszynski2011}, compared to the 79,200 \sargs\ in the LMC \citep{Soszynski09}. Figure\,\ref{fig:PL_AllSequences} shows the Period-Wesenheit relations of the OGLE long-period variables, which exhibit multi-periodic variability and populate multiple period-luminosity sequences of both radial and non-radial modes, labeled  A$'$, A, B, and so on by \citet{Wood1999,Wray2004,Wood15}. These PL sequences have been studied intensively by many authors \citep[e.g.,][]{Cioni2000,Ita2004}, and with occasionally different nomenclature. Specifically, the sequences A \& B used here follow the nomenclature by Wood and were labeled R3 and R2, respectively, in \citep{RedVarsIKiss2003,RedVarsIIKiss2004}. Additionally, \citet{Soszynski2004} distinguished eight sequences of SARGs, four above and four below the TRGB. Our A \& B sequences correspond to the combinations of their $a3$ and $b3$, and $b2$ and $a2$ sequences, respectively. The fact that individual PL-sequences of LPVs show the TRGB feature was first pointed out by \citet{Ita2002}.

\begin{figure*}[ht!]
\includegraphics[width=1\textwidth]{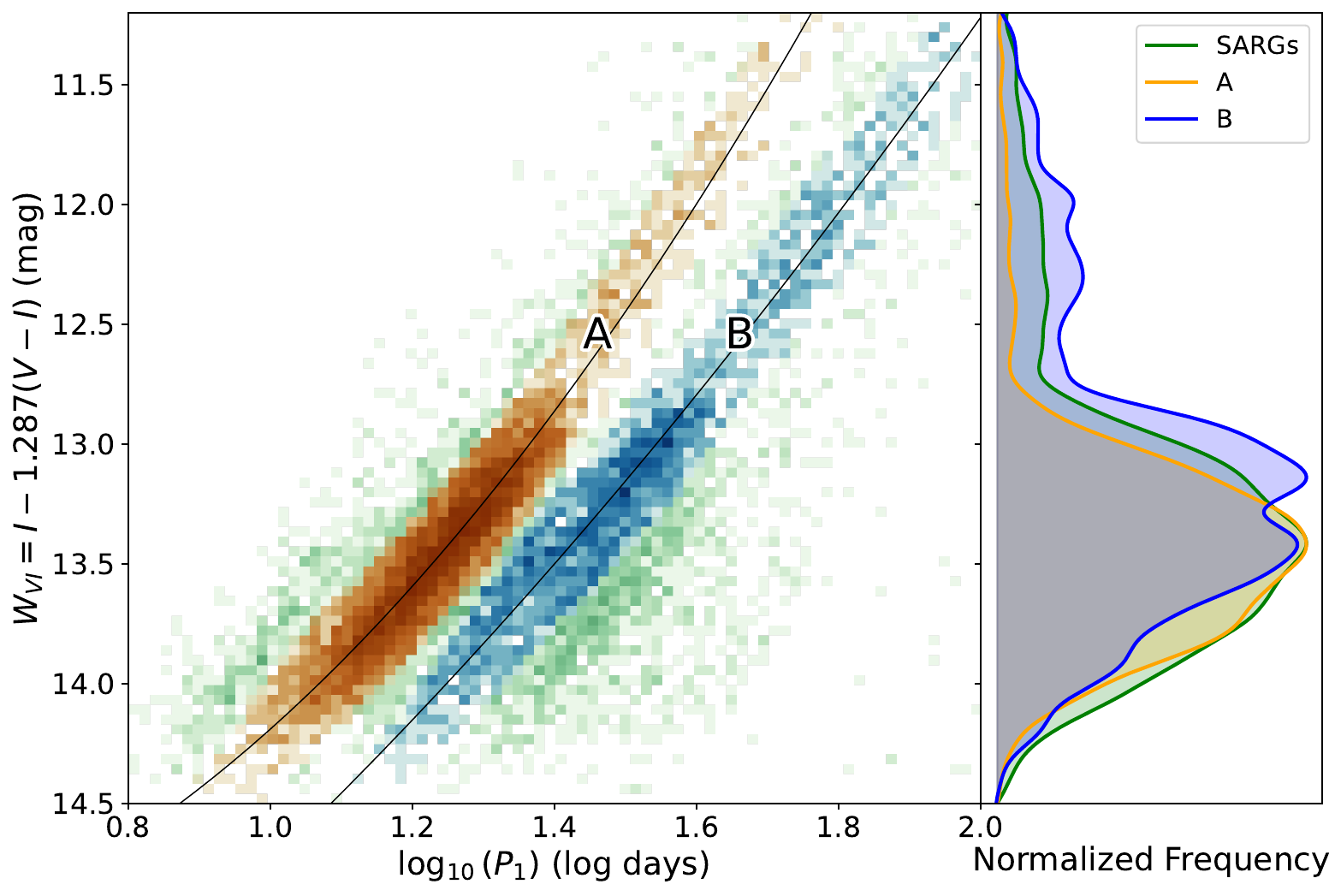}
    \caption{\textit{Left:} Density plot (log-scaled star count) of period-luminosity relations for SARGs. The stars shown are identical to the ones in Fig.\,\ref{fig:PL_AllSequences}. The two sequences selected for this work, $A$, and $B$ are shown using orange and blue colors against the background of all SARGs shown in green. They were selected using $3\sigma$ ranges around the polynomial regressions shown as thin black lines, leaving some stars in between the selected areas that could not be uniquely associated to either sequence.  \textit{Right:} Normalized luminosity functions for the three SARG samples. GLOESS smoothing with $\sigma_s = 0.11$ mag has been applied.}
    \label{fig:PL}
\end{figure*}

We selected two further subsamples, \Aseq\ and \Bseq, according to the period-luminosity relation of their dominant pulsation period, $P_1$, in analogy to A24. Specifically, we used second order polynomial fits to period-Wesenheit relations computed using $W_{VI} = I-1.287(V-I)$, which is reddening-free by construction \citep{madore_pl_1982} assuming $R_V = 2.7$ and a typical color of stars near the RGB tip \citep{Anderson2022}. Wesenheit magnitudes were only used to select samples, cf. Sect.\,\ref{sec:photometry}. Three$-\sigma$ clipping was applied to avoid overlap between both sequences, which results in a small number of stars that cannot be uniquely associated with either sequence falling in between them. To avoid contamination for the \Bseq\ sample, we restricted the selection using a polygon designed to exclude a longer-period, lower-luminosity sequence situated below the \Bseq, which corresponds to the b$_1$ sequence in \cite{Soszynski2004} and does not reach the TRGB. Exclusion of these stars has therefore no impact on our TRGB measurement. The polynomial fits are for \Aseq: $ -1.68(\log P_1)^2 + 0.71 (\log P_1) + 15.16$ mag with a dispersion of $0.12$\,mag and for \Bseq: $ -0.68(\log P_1)^2 - 1.48 (\log P_1) + 16.91$ mag with a dispersion of $0.14$\,mag. The selected sequences can be seen in Figure~\ref{fig:PL} along with their $W_{VI}$ luminosity functions and the collected information for each sample is listed in Table~\ref{tab:color}. 

\begin{figure}
    \includegraphics[width=0.5\textwidth]{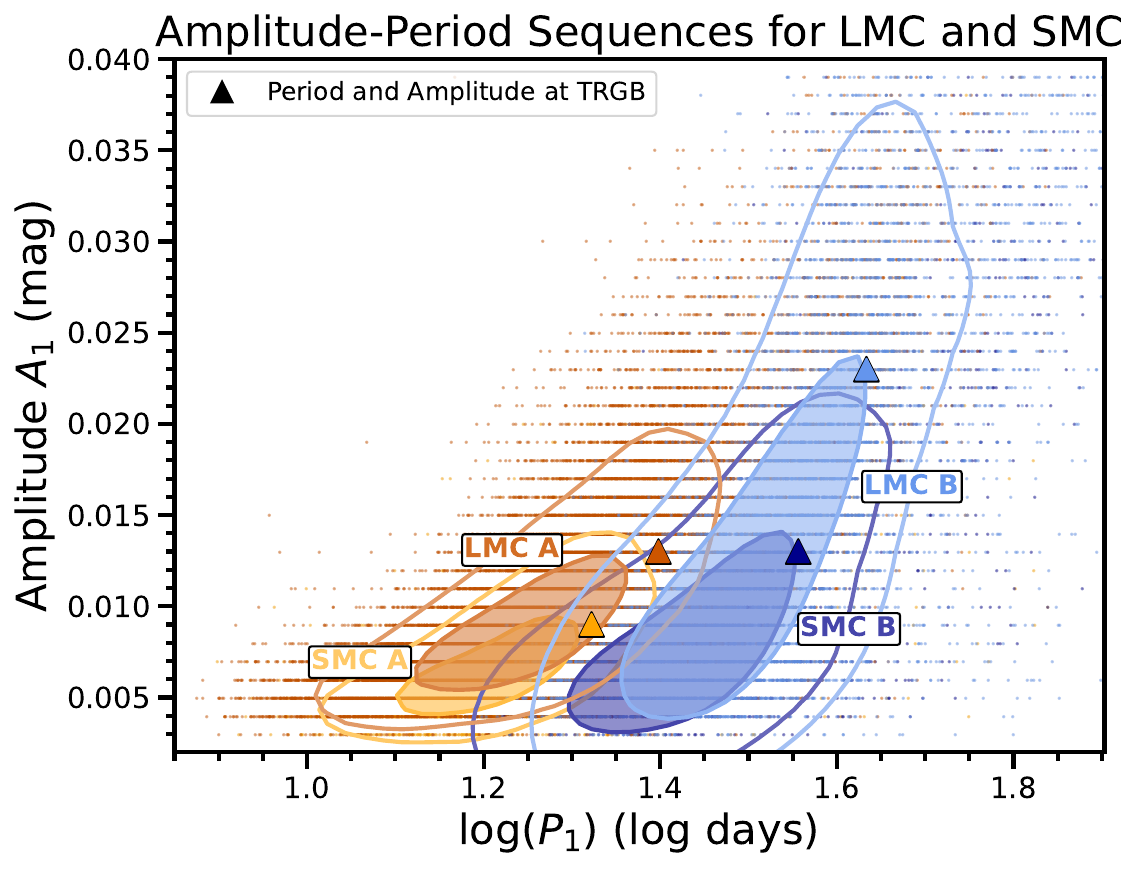}
    \caption{Period-amplitude relation of SARGs in the SMC and LMC. Only the primary modes are considered. Triangles indicate the median amplitudes and periods measured within 0.1 mag of the TRGB feature determined below. Sequence B exhibits a larger range in amplitude than Sequence A. However, Sequence B cannot be isolated using a cut in amplitude alone. LMC SARGs exhibit larger amplitudes in the LMC than in the SMC, especially on the B-sequence.}
    \label{fig:AmplitudePeriod}
\end{figure}

Figure\,\ref{fig:AmplitudePeriod} shows the period-amplitude relations for \sargs\ on the A and B-sequences, which feature oscillation amplitudes ($A_1$ from the \ogle\ catalog by \citealt{Soszynski2004,Soszynski2011}) $\sim 0.01-0.02$ mag near the TRGB. Interestingly, SMC \Bseq\ stars exhibit smaller amplitudes near the TRGB than LMC \Bseq\ stars. This is further discussed in Section\,\ref{sec:spectra} below.

\noindent
\begin{table*}[]
    \caption{Properties of the RG samples considered. \label{tab:color}}
    \hskip -4cm
    \begin{tabular}{@{}l@{\hskip 2mm}c@{\hskip 2mm}c@{\hskip 2mm}c@{\hskip 2mm}c@{\hskip 2mm}c@{\hskip 2mm}c@{\hskip 2mm}c@{\hskip 2mm}c@{\hskip 2mm}c@{\hskip 2mm}c@{\hskip 2mm}c@{\hskip 2mm}c@{\hskip 2mm}c@{\hskip 2mm}c@{}}
    \toprule
        & \multicolumn{2}{c}{$\langle(V-I)_0\rangle$} & \multicolumn{2}{c}{$\langle P_1 \rangle$} & \multicolumn{2}{c}{$\langle A_1 \rangle$} & \multicolumn{2}{c}{$[\mathrm{Fe/H}]$} & \multicolumn{2}{c}{$\mathrm{Mass}$} &  \multicolumn{2}{c}{$\mathrm{Age}$} & \multicolumn{2}{c}{Number of Stars}  \\
        & \multicolumn{2}{c}{(mag)} & \multicolumn{2}{c}{(d)} & \multicolumn{2}{c}{(mmag)} & \multicolumn{2}{c}{} & \multicolumn{2}{c}{($M_\odot$)}  & \multicolumn{2}{c}{(Gyr)} & \\
        
         Sample & LMC & SMC & LMC & SMC & LMC & SMC & LMC & SMC & LMC & SMC & LMC & SMC & LMC & SMC \\
\midrule
A & 1.78 (0.30) & 1.52 (0.15) & 25 & 21 & 13 & 9 & -0.63 (0.17) & -0.95 (0.25) & 1.2 & 1.2 & 3.5 & 2.9 &  20 470&5 468\\
B & 1.84 (0.38) & 1.57 (0.18) & 43 & 36 & 23 & 13 & -0.77 (0.35) & -1.09 (0.29) & 0.9 & 1.0 & 6.3 & 4.1 & 9 164&1 943\\
\sargs\ & 1.80 (0.37) & 1.55 (0.18) & 39 & 29 & 19 & 12 & -0.70 (0.32) & -1.00 (0.28) & 1.0 & 1.1 & 4.6 & 3.2 & 40 185&11 584 \\
All Stars & 1.78 (0.42) & 1.53 (0.23) & - & - & - & - & -0.68 (0.35) & -1.00 (0.31) & 1.1 & 1.1 & 4.3 & 3.2 &  140 774&45 821\\
\bottomrule
    \end{tabular}
    \tablecomments{Sample properties are reported for both the LMC and the SMC. Median color $\langle (V-I)_0 \rangle$ with 16th-84th percentile range in parentheses, median period $\langle P_1 \rangle$, median amplitude $\langle A_1 \rangle$, mean metallicity [Fe/H] with 16th-84th percentile range in parentheses  \citep{GarciaPerez2016}, median mass,  median age \citep{Povick2023, Povick2023SMC}, and the number of stars per sample. Medians and 16th-84th percentile ranges for color and metallicity are determined using RGs within 0.1 mag of the TRGB. Rather few stars have known [Fe/H]. In the LMC, \AllStars: 362, \Aseq: 160, \Bseq: 65, and \sargs\: 323. In the SMC: \AllStars: 207, \Aseq: 62, \Bseq: 46, and \sargs\: 190. 
    Ages and masses of stars whose uncertainties did not exceed $3$ Gyr and $3\,M_\odot$, respectively, were considered to determine the listed averages. 
    The \Bseq\ is consistently older than the \Aseq\ in both galaxies. Metallicity from \cite{GarciaPerez2016} is lower in the SMC across all sequences, supporting findings from A24. SMC samples are younger than LMC samples according to ages determined by \citet{Povick2023,Povick2023SMC}.}
\end{table*}

\subsection{Photometric data collected\label{sec:photometry}}
\begin{figure*}
    \includegraphics[width=1\textwidth]{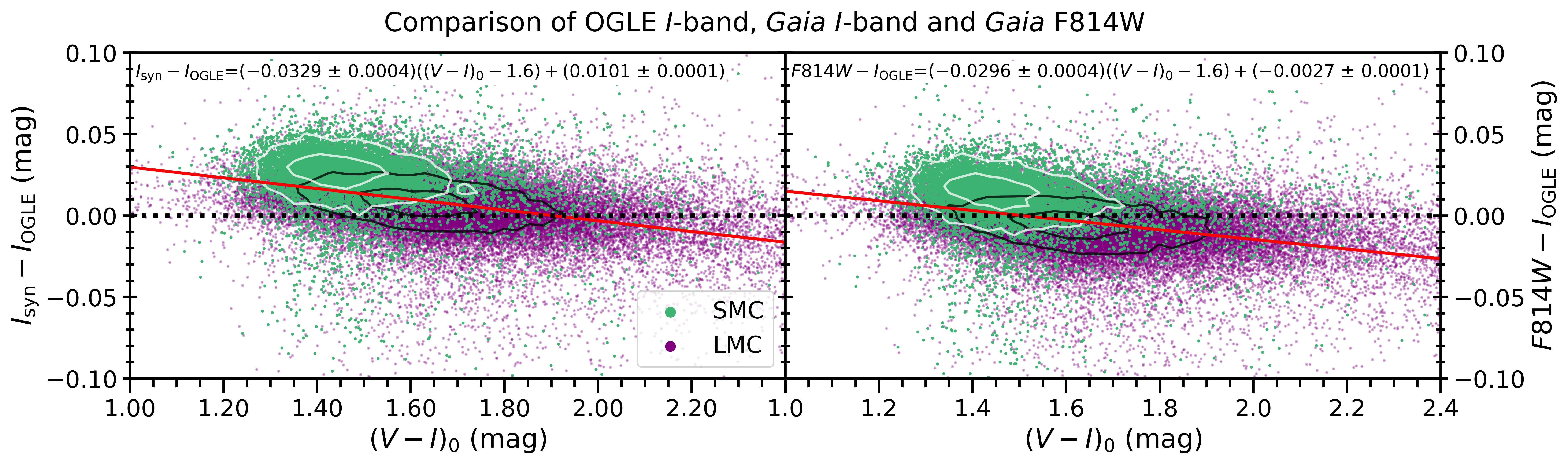}
    \caption{\textit{Left:} The difference between \gaia\ synthetic Cousins $I$-band and OGLE-III $I$-band magnitudes for SMC (green) and LMC (purple) \sargs\ against de-reddened OGLE-III color. \textit{Right:} Difference between \gaia\ synthetic HST ACS F814W and OGLE-III $I$-band magnitudes. The photometric systems are similar, but not identical, and the difference is greater in the SMC than in the LMC. We find a decreasing trend of magnitude difference $\left(I_{\mathrm{syn}}-I_{\mathrm{OGLE}}\right)$ with color. The y-axis is restricted to differences between -0.1 and 0.1 for clarity. Also included are the 95th and 99th percentile contours for the SMC (white) and LMC \sargs\ (black) and a combined linear fit (in red) based on all 51,764 \sargs\ in both galaxies.}
    \label{fig:GaiaVsOGLEIband}
\end{figure*}

We computed mean magnitudes as the straight average of $V$ and $I-$band time-series photometry from the SMC \citep{Soszynski2011} and LMC OGLE-III studies \citep{Soszynski09}. For non-variable stars we used the OGLE-III photometric maps of the SMC \citep{Udalski2009} and LMC \citep{Udalski08}, removing duplicate stars in the photometric maps within $0.2"$. The mean time-series magnitudes and the photometric maps are fully consistent with each other, with a mean difference of $0.00002$\,mag.

We cross-matched the OGLE stars with \gaia\ DR3 \citep{GDR3_summary} to collect astrometry and photometry in \gaia\ $G_{RP}-$band and synthetic photometry from the \texttt{gaiadr3.synthetic\_photometry\_gsp} table that provides \emph{HST} ACS/WFC $F814W$ band and \emph{Cousins} $I-$band, among others, in addition to photometric quality indicators. We applied very loose constraints on the color-magnitude diagram, as well as proper motion and quality selections, following A24. Our \gaia\ query for the SMC was centered on the SMC central region \citep{Graczyk2020}:

\begin{quote}
\small
\tt
SELECT * FROM gaiadr3.gaia\_source as GS \\
INNER JOIN gaiadr3.synthetic\_photometry\_gspc as S \\
ON S.source\_id = GS.source\_id\\
WHERE CONTAINS(POINT('ICRS',GS.ra,GS.dec),\\CIRCLE('ICRS',$12.5$,$-73.1$,$3.3$))=1\\
AND S.i\_jkc\_mag $>13$\\ AND S.i\_jkc\_mag $<17.5$ \\
AND (S.v\_jkc\_mag - S.i\_jkc\_mag) $> 1.0$\\ AND (S.v\_jkc\_mag - S.i\_jkc\_mag) $< 3.5$
\end{quote}

Several additional quality cuts similar to those recommended by \cite{Riello2021} and \cite{Montegriffo2022} were applied to remove stars likely affected by blending or poor photometry which amounted to $\sim$30-36\% of stars depending on the sample. These cuts are detailed in Table 3 of Appendix A in A24, and include cuts on the following parameters: \texttt{ipd\_frac\_multi\_peak}, \texttt{ipd\_frac\_odd\_win},  $C^*$, and $\beta$ \citep{Riello2021}.

\label{sec:Ioglevgaia}
$I-$band photometry from OGLE-III, \gaia's synthetic Cousins $I-$band, and \gaia's synthetic F814W  band are all similar, although slightly different. We investigated differences between OGLE $I-$band and \gaia's synthetic Cousins $I-band$ and F814W for all SMC and LMC \sargs\ and found a significant trend with color shown in Figure~\ref{fig:GaiaVsOGLEIband}. Fitting the trends with straight lines, we obtained $F814W-I_{\text{OGLE }}=(-0.0296 \pm 0.0004 )\left((V-I)_0-1.6\right)+(-0.0027 \pm 0.0001)$ mag with an rms scatter of approximately 0.026 mag, in line with the dispersion of $\sigma \approx 0.02$ mag found for the validation of \gaia\ synthetic photometry based on globular clusters \citep{Montegriffo2022}. Despite the scatter, the color-dependence is highly significant and should be accounted for when mixing photometric systems. Interestingly, we find that the SMC  is offset on average by $\sim 0.02$\,mag relative to the LMC in both passband comparisons. Although the contours of both galaxies have high overlap, we note that such an offset could be caused by increased blending in the \ogle\ photometry at the slightly larger distance to the SMC. To avoid issues related to mixed photometric systems, we primarily considered TRGB measurements based on \gaia's synthetic F814W photometry, although we also reported measurements based on OGLE $I-$band, \gaia\ Cousins $I-$band, and \gaia\ $G_{Rp}$ spectrophotometric magnitudes. Additionally, all comparisons between LMC and SMC are reported using identical photometric systems.

Suspected foreground stars were removed if they met the following criteria: the star had high quality astrometry (\emph{RUWE}$\, < 1.4$), and either the star has a parallax satisfying $\varpi - \sigma_\varpi > 1/62.4\,\text{kpc}$ with a signal-to-noise ratio of $\varpi / \sigma_\varpi > 5$, or the star was outside the proper motion ellipse: $\left(- 2.12\cdot(\mu_\delta + 1.23)\right)^2 + \left( 1.6\cdot(\mu_\alpha^* - 0.66)\right)^2 < 1$ \citep{GaiaEDR3_LMC}.

\subsection{Additional information considered\label{sec:spectra}}

We further collected information on iron abundance ([Fe/H]) and stellar ages based on infrared spectroscopy from the Apache Point Observatory Galactic Evolution Experiment using the APOGEE Stellar Parameters and Chemical Abundances Pipeline \citep{GarciaPerez2016,Jonsson2020} and specific studies of RGs in the Magellanic Clouds \citep{Povick2023, Povick2023SMC}. Table~\ref{tab:color} provides an overview of the RG populations in the SMC and LMC based on this information. For each RG sample, we determined the median color, period, amplitude, [Fe/H], mass, and age of all stars within $0.1$\,mag of the TRGB. We note that [Fe/H], age, and mass were derived from much fewer stars than the other parameters due to the smaller spectroscopic dataset.

\begin{figure*}[]
\includegraphics[width=1.0\textwidth]{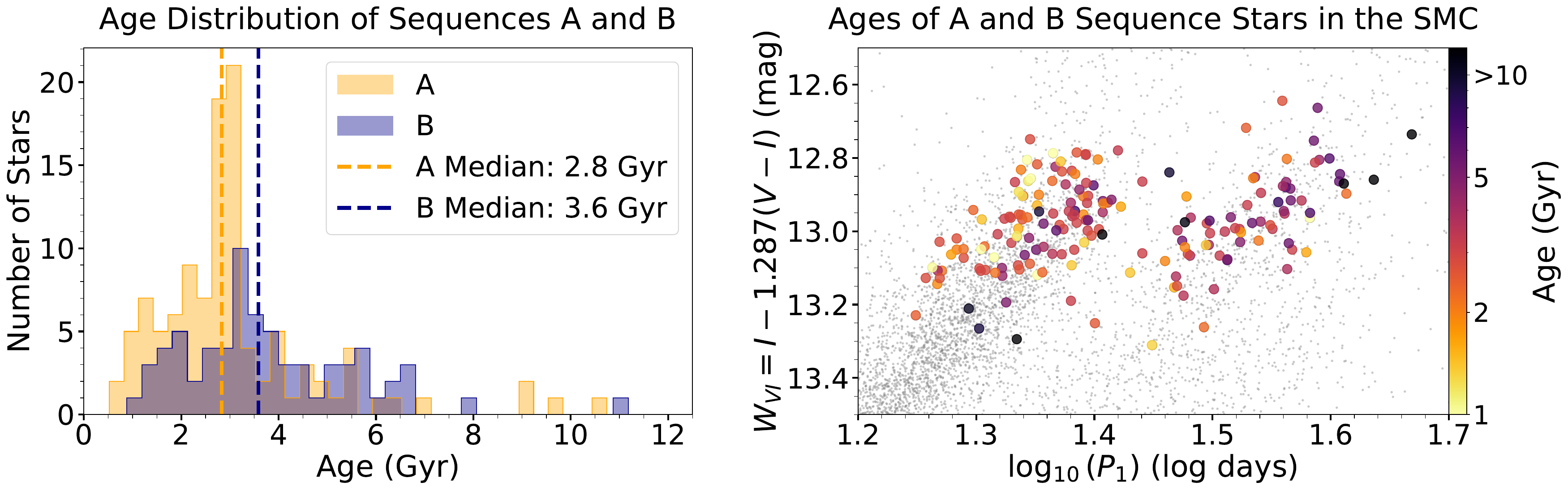}
    \caption{Ages of \Aseq\ and \Bseq\ stars  derived from APOGEE spectra by \citet{Povick2023SMC}. \textit{Left:} Histogram showing ages of stars on Sequences A and B. As reported in the LMC (A24), stars on Sequence B are on average older than stars on Sequence A. The medians reported here differ slightly from  Table~\ref{tab:color}, since we here did not restrict to stars within 0.1 mag of the TRGB. \textit{Right:} Period-Wesenheit relations of Sequences A and B color-coded by age. As observed in the LMC, the short period edge of the \Aseq\ seems to be younger than the long period edge, and there is an age gradient across both sequences individually. Stellar ages with errors exceeding $> 100\%$ are not shown (A: 8 stars, B: 3 stars).}
    \label{fig:Age}
\end{figure*}

As expected, we see that the SMC stars are slightly more metal-poor than the LMC, with the LMC stars near [Fe/H] = -0.7 dex and the SMC near [Fe/H] = -1.0 \citep{GarciaPerez2016}. Additionally, Fig.~\ref{fig:Age} shows ages from \citet{Povick2023SMC} and reveals \Bseq\ RGs to be older than \Aseq\ RGs in the SMC, as A24 found in the LMC. This confirms the evolutionary scenario of \sargs\ proposed by \citet{Wood15}, which describes the P-L sequences of long-period variables as an evolutionary sequence. Interestingly, however, the SMC RGs are overall younger than the RGs in the LMC. 
Further support for the evolutionary scenario of the LPV sequences comes from the mass differences between \Aseq\ and \Bseq\ stars, which are $33\%$ in the LMC and $16\%$ in the SMC, respectively, and agree well with the expectation based on pulsation models \citep[$26\%$]{Wood15}.

\begin{figure}[]
    \includegraphics[width=0.5\textwidth]{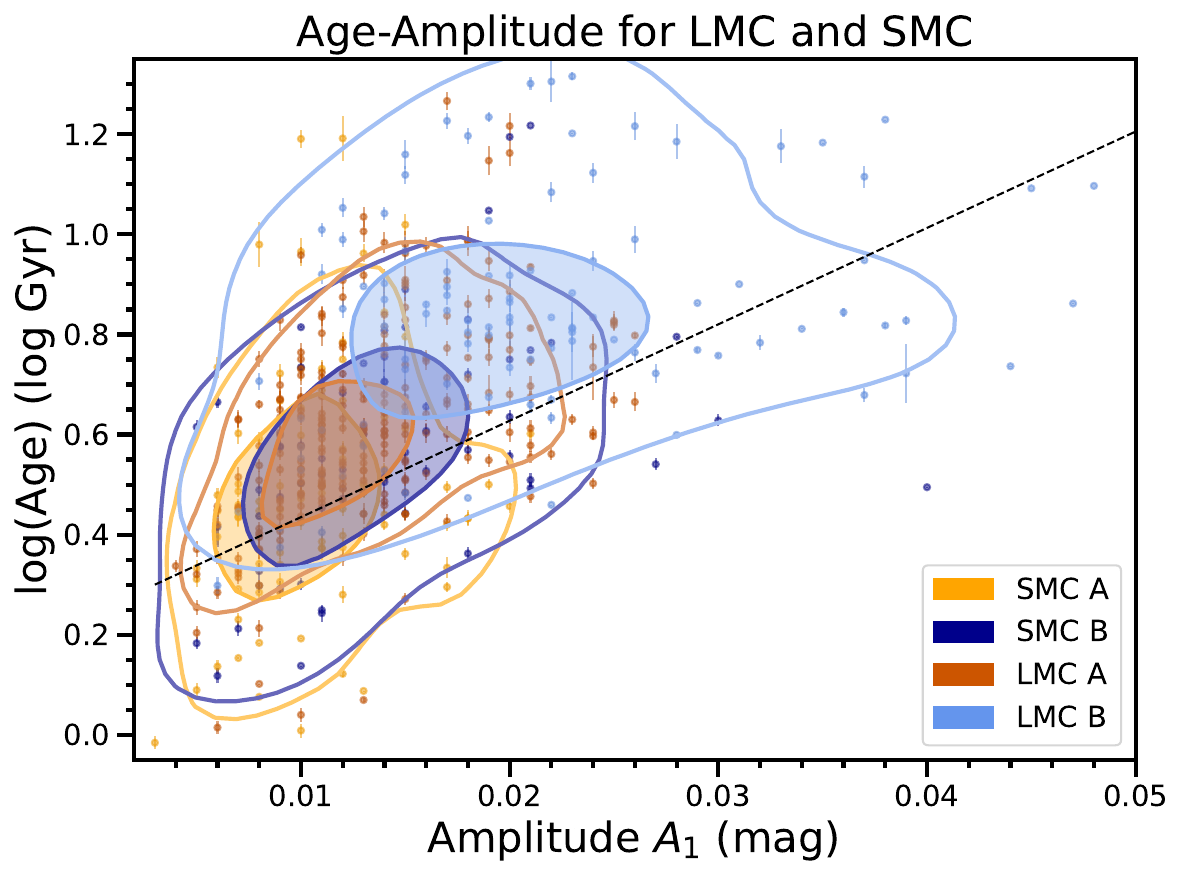}
    \caption{Age-amplitude relations of \sargs\ below the TRGB in the SMC and LMC. Ages were determined using APOGEE spectra by \citet{Povick2023} and \citet{Povick2023SMC}. The contours and a linear fit of all points (black dashed line) show that larger pulsation amplitudes correspond to older stars. 
    A fit to all points yields $\log(\mathrm{age [Gyr]})=(19.3 \pm 0.4) \cdot A_1  + (0.24 \pm 0.02))$. } 
    \label{fig:AgeAmplitude}
\end{figure}

Figure~\ref{fig:AgeAmplitude} shows that older stars have larger amplitudes in both the \Aseq\ and \Bseq\ samples, and that the older LMC stars have larger amplitudes on both sequences than the younger SMC stars. It thus appears that the variance of RG magnitudes in an extragalactic context could usefully identify the older, larger-amplitude variability of the older RG populations, which are particularly desirable for extragalactic TRGB measurements. We note that the correlation between amplitude and age is not a selection effect related to photometric incompleteness since the smaller amplitudes are found among the fainter stars in the SMC.

\subsection{Reddening\label{sec:reddening}}
All stars are de-reddened ($\textit{m}_{\lambda,0} = \textit{m}_\lambda  - R_\lambda E(V-I)$) using the reddening map for the Magellanic system based on OGLE-IV Red Clump stars \citep{Skowron2021ApJS} and $R_{\mathrm{I}}=1.290$, $R_{\mathrm{ACS,F814W}}=1.248$,  $R_{G_{RP}}=1.319$ calculated using \emph{pysynphot} \citep{pysynphot}, assuming the spectrum of a typical RG star near the tip \citep{Anderson2022} and the recommended $R_{V} = 2.7$ value for the SMC from \cite{Bouchet1985} and \cite{Gordon2003} who derived it using O-B stars. As a check, we also considered $R_V = 3.3$ in Section~\ref{sec:mudiff}, which yields $R_I=1.461$ (as in A24). We removed stars with $E(V-I) \geq 0.2$ mag to reduce the impact of reddening law uncertainties, leading to a mean $E(V-I)$ of $0.05$\,mag.

We averaged the statistical uncertainties from reddening maps ($\sigma_1$ and $\sigma_2$) for each  $E(V-I)$ value obtained from \cite{Skowron2021ApJS} and combined them in quadrature with the photometric statistical uncertainties. The reddening uncertainties clearly dominate the combined statistical photometric uncertainties, whose average turned out to be $0.06$\,mag for \sargs. These were used in the Monte Carlo method to determine a final statistical uncertainty for the TRGB magnitude.

\subsection{Determining the TRGB Magnitude}\label{sec:MC}

The process for determining the TRGB follows the methodology detailed by A24 and adapted from \cite{Hatt17}. First, a smoothed luminosity function (LF) is obtained by binning the dereddened $I-$band magnitudes and in turn smoothing the LF using a Gaussian-windowed LOcal regrESSion (GLOESS) algorithm that depends on the smoothing parameter $\sigma_s$ \citep{Persson04}. 
The measured TRGB magnitude, \mtrgb, corresponds to the inflection point of this smoothed LF and is determined by the maximum of an unweighted [-1, 0, +1] Sobel filter response curve. While several recent studies adopted a weighted Sobel filter response curves to determine \mtrgb\ \citep{Scolnic2023, Wu2022, Li2023, Hoyt2023}, we prefer an unweighted Sobel filter edge detection response (EDR) because weighting introduces a systematic on \mtrgb\ that depends on the properties (specifically contrast) of the LF, which can only be assessed a posteriori (A24). 

Uncertainties are determined through a Monte Carlo simulation that remeasures the TRGB after sampling stellar magnitudes from Gaussian distributions, using the photometric and reddening errors. The Monte Carlo was iterated 1000 times to extract a mean TRGB and standard deviation for the range $\sigma_s \in [0.01,0.50]$. 

A24 pointed out the importance of bias introduced by smoothing, which can impact the value of \mtrgb\ depending on the shape of the observed LF. Following A24, we measured the global \mtrgb\ value using the range of \sigs\ values where \mt\ remains insensitive to the smoothing parameter as measured by the derivative, $|\mathrm{dm}_{\mathrm{TRGB}}/\mathrm{d}\sigma_s|$. Specifically, we considered the lowest, continuous \sigs\ range satisfying $|\mathrm{dm}_{\mathrm{TRGB}}/\mathrm{d}\sigma_s| \leq 0.1$. The reported values of \mtrgb\ are the median of the \mtrgb$(\sigma_s)$ values satisfying these criteria.

\begin{table*}[ht!]
\caption{Uncertainty budget for TRGB measurements}
\hspace{-1.5cm}
\begin{tabular}{lllcc}
   \toprule 
   Uncertainty  & Includes / based on & \\
   \midrule 
   \multicolumn{3}{l}{Estimation of statistical uncertainty} \\
   \midrule
    \sigphot\ & \multicolumn{4}{l}{Average: 0.062\,mag. Computed as quadratic sum of:} \\
    & \multicolumn{4}{l}{Photometric uncertainties  \citep{Udalski2009,Soszynski2011,Montegriffo2022}} \\
    & \multicolumn{4}{l}{$R_I \times \sigma_{EVI,\mathrm{stat}}$, with $\sigma_{EVI,\mathrm{stat}}$ the average of $\sigma_1$, $\sigma_2$ from \cite{Skowron2021ApJS}} \\
    & \multicolumn{4}{l}{$R_I \times \sigma_{EVI,\mathrm{sys}}$,  with $\sigma_{EVI,\mathrm{sys}} = 0.014$\,mag, the systematic uncertainty from \cite{Skowron2021ApJS}} \\
    & \multicolumn{4}{l}{\gaia\ GSPC: standardization uncertainties from \cite{Montegriffo2022}}\\
    \sigmc & \multicolumn{4}{l}{Dispersion of 1000 MC resamples per \sigs\ value} \\
    \textbf{\sigt} &\multicolumn{4}{l}{\textbf{Total statistical uncertainty on \mtrgb}: median \sigmc\ across \sigs\ range, where $\vert dm_I / d\sigma_s\vert \le 0.1$} \\
        &\multicolumn{4}{l}{In case of \Mtrgb\ also considers statistical uncertainty of geometric distance} \\
   \midrule
    \multicolumn{4}{l}{Systematic uncertainties of \mt, combines bin size \& phase, choice of $R_V$, and algorithmic aspects} \\
    \midrule
   \multicolumn{2}{l}{\Aseq} & & & 0.015\\
   \multicolumn{2}{l}{\Bseq} & & & 0.008 \\
   \multicolumn{2}{l}{\varstars} & & & 0.006 \\
   \multicolumn{2}{l}{ \AllStars} & & & 0.008 \\
    \midrule
   \multicolumn{3}{l}{Distance-related uncertainties} & (stat.) & (syst.)\\
   \midrule
   \multicolumn{2}{l}{SMC DEB distance} & from \cite{Graczyk2020}: $\mu = 18.977$\,mag &  $0.016$ & $0.028$ \\
   \multicolumn{2}{l}{LMC DEB distance} & from \citet{Pietrzynski19}: $\mu = 18.477$\,mag &  $0.004$ & $0.026$ \\
   \multicolumn{2}{l}{DEB distance difference} & from \cite{Graczyk2020}: $\Delta \mu_{\mathrm{SMC-LMC,DEB}} = 0.500$\,mag & \multicolumn{2}{c}{$0.017^\dagger$} \\
   \bottomrule
   \end{tabular}
\tablecomments{The top part of the table describes the composition of statistical errors \sigt\ reported in Tab.\,\ref{tab:AllTRGBS}. \sigmc\ is determined per smoothing value ($\sigma_s$) via Monte Carlo resampling. \sigphot\ is dominated by the reddening correction uncertainties and differs per star. The middle part reports systematic uncertainties (in mag) associated with measuring \mtrgb\ for each of the samples considered. These include bin size and phase variations ($0.004$\,mag), half the difference of \mt\ when assuming $R_V=2.7$ vs $R_V=3.3$, and the estimate of the method's accuracy determined by simulations following A24, adapted to the SMC's LF shapes ($0.010$\,mag for \Aseq, $0.005$\,mag for the other samples). The bottom part lists statistical and systematic uncertainties (in mag) of the DEB distances. $^\dagger$: combined statistical and systematic uncertainty, applies to distance modulus difference (Sect.\,\ref{sec:mudiff}).} \label{tab:errors}
\end{table*}

\subsection{Overview of statistical and systematic uncertainties}\label{sec:sys_unc}
Table~\ref{tab:errors} summarizes the systematic uncertainties applicable to our results, including uncertainties related to photometric systems, dereddening, metallicity corrections, and the algorithmic implementation of TRGB determination. We further note the importance of measuring \mtrgb\ using consistent and properly standardized methods to avoid bias. As a reminder, we used an unbiased [-1,0,1] Sobel filter for edge detection to avoid correlations between the measured \mtrgb\ and tip contrast.

We report as statistical uncertainties on \mtrgb, \sigt, the median of the standard deviations across the range of $\sigma_s$ values, where $|\mathrm{dm}_{\mathrm{TRGB}}/\mathrm{d}\sigma_s| \leq 0.1$. The Monte Carlo resampling considers the total photometric uncertainty for each star, \sigphot, obtained by the squared sum of reported photometric uncertainties, the uncertainty of photometric standardization (in the case of \gaia\ synthetic photometry), and reddening-related uncertainties, which dominate the overall error budget. Absolute magnitudes further consider the statistical uncertainties of the geometric distances.

Differences between photometric systems were considered in Section~\ref{sec:photometry} and can be applied if needed. However, we considered only direct comparisons between identical photometric bands.

Systematic uncertainties on \mtrgb, \sigs, include contributions from bin size and phase variations (estimated following A24), the choice of the reddening law (cf. Section~\ref{sec:reddening}), and simulation-based bias estimates following Appendix B in A24, adapted here to match the LF shapes in the SMC. We note that the systematic error of the red clump color excesses of $0.014$\,mag \citep{Skowron2021ApJS} is already included in \sigphot. Absolute magnitudes further consider the systematic uncertainties of the geometric distances.

\subsection{Absolute magnitudes and relative distances}

We determined absolute TRGB magnitudes, \Ml, using the measured apparent magnitudes, \mtrgb, and the known distance modulus of the SMC, $\mu_{\mathrm{SMC}} = 18.977 \pm 0.016 \,\mathrm{(stat.)} \pm 0.028 \,\mathrm{(sys.)}$ mag, from \cite{Graczyk2020}. In the case of the LMC, we use  $\mu_{\mathrm{LMC}} = 18.477 \pm 0.004\,\mathrm{(stat.)} \pm 0.026\,\mathrm{(sys.)}$ mag from \cite{Pietrzynski19}. 

We further determined the relative distance modulus, $\Delta \mu_{\mathrm{SMC-LMC}}$, by considering the difference of the apparent magnitudes of the RGB tips in both galaxies, $\Delta \mu_{\mathrm{SMC-LMC}} = m_{\lambda\mathrm{, SMC}} - m_{\lambda\mathrm{, LMC}}$. We compared these distance differences to the equivalent number determined using the geometric distances of the detached eclipsing binaries ($\Delta \mu_{\mathrm{SMC-LMC,DEB}} = 0.500 \pm 0.017$ mag) based on 15 systems in the SMC and 20 in the LMC \citep{Graczyk2020,Pietrzynski19}. We note that the uncertainty on $\Delta \mu_{\mathrm{SMC-LMC,DEB}}$ is smaller than the squared sum of  uncertainties for both galaxies due to shared systematics.

\section{Results}\label{sec:results}

A24 recently showed that virtually all red giant stars near the RGB tip in the LMC are \sargs. Figure~\ref{fig:CMD} shows the analogous behavior for red giants in the SMC: nearly 100\% of SMC stars near the RGB tip inside the OGLE-III footprint are \sargs. Given the agreement between LMC and SMC in this regard, it seems clear that this level of variability is an astrophysical property of stars near the helium flash.

Figure\,\ref{fig:CMDs} shows \ogle-III color-magnitude diagrams for the three SARG samples and shows that color-magnitude cuts cannot distinguish the stars on the {\tt A-} and \Bseq. From Fig.\,\ref{fig:AmplitudePeriod}, it is clear that the periods of the A \& B-sequence \sargs\ are systematically shorter in the SMC compared to the LMC in addition to featuring smaller amplitudes. The following first measures the SMC TRGB magnitudes and then proceeds to exploit variability features (periods and amplitudes) to investigate the impact of astrophysical differences on TRGB measurements.
 
\subsection{Apparent TRGB magnitudes in the SMC\label{sec:mtrgb}}
\begin{figure*}[ht!]
    \centering
    \includegraphics[width=0.75\textwidth]{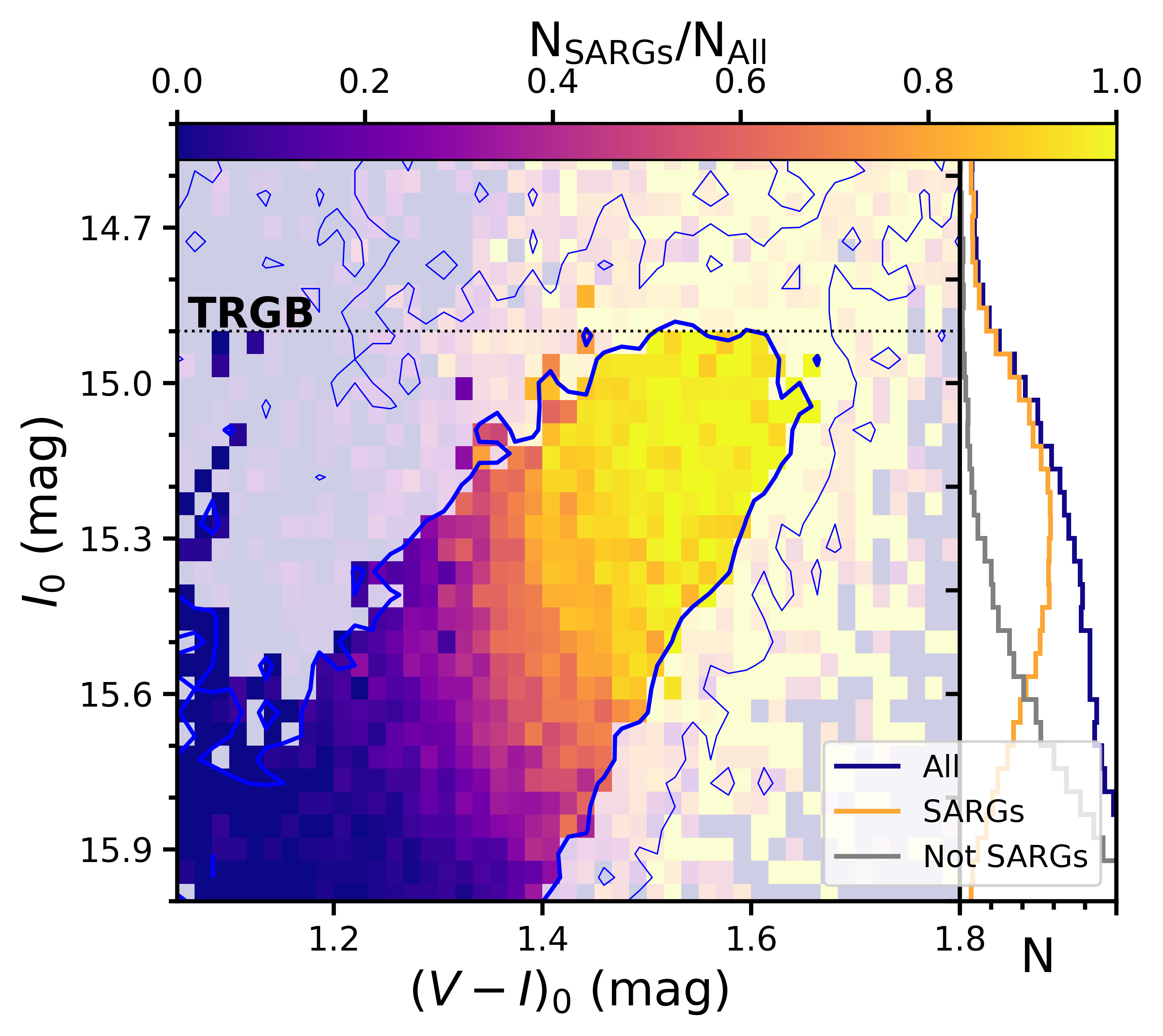}
    \caption{All RGs near the TRGB feature in the SMC are SARGs. \textit{Left:} CMD showing the percentage of red giant stars in the SMC that exhibit variability and are classified by OGLE-III as a SARG \citep{Soszynski2011}. The smoothed contours encapsulate bins with more than 5 stars (thin) and 20 stars (thick). Similar to Fig. 1 in A24 with the LMC, the fraction of RG stars being variable increases to close to 100\% near the tip (indicated by a dashed line). \textit{Right:} The luminosity functions of all RG stars, \sargs\, and not \sargs.}
    \label{fig:CMD}
\end{figure*}

\begin{figure*}[ht!]
    \centering
    \includegraphics[width=0.8\textwidth]{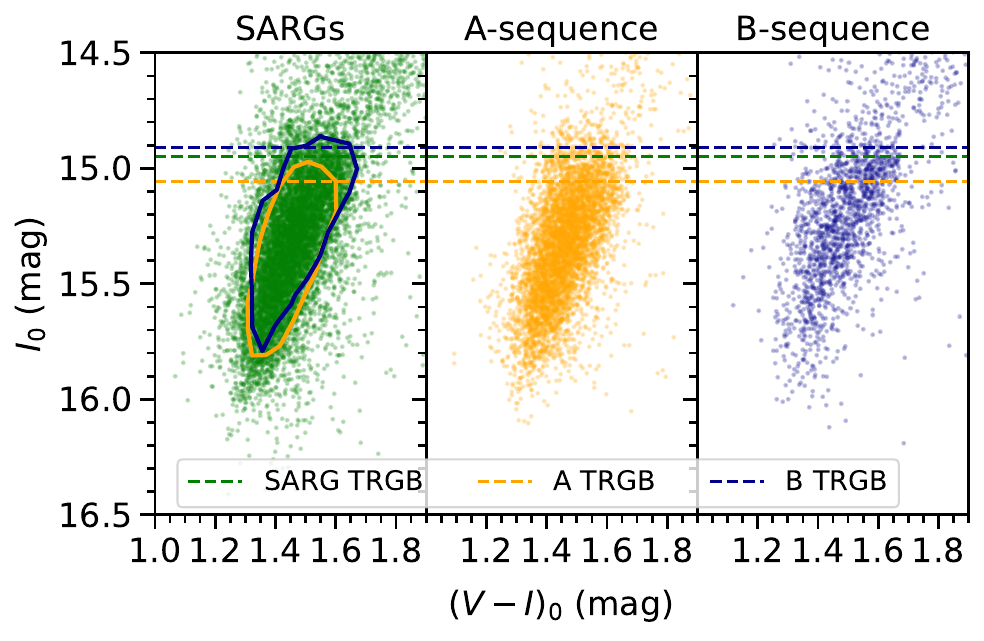}
    \caption{Color-magnitude diagrams of SARG samples in the SMC based on \ogle\ photometry. {\it Left:} The green points represent all SARGs, of which the \Aseq\ and \Bseq\ samples are subselections. Dashed horizontal lines indicate the results obtained for each sample. Contours for \Aseq\ (yellow) and \Bseq\ stars (blue) show that color-magnitude selections do not allow to distinguish the samples. {\it Center:} \Aseq\ stars. {\it Right:} \Bseq\ stars.}
    \label{fig:CMDs}
\end{figure*}

\begin{figure*}[ht!]
\includegraphics[width=1\textwidth]{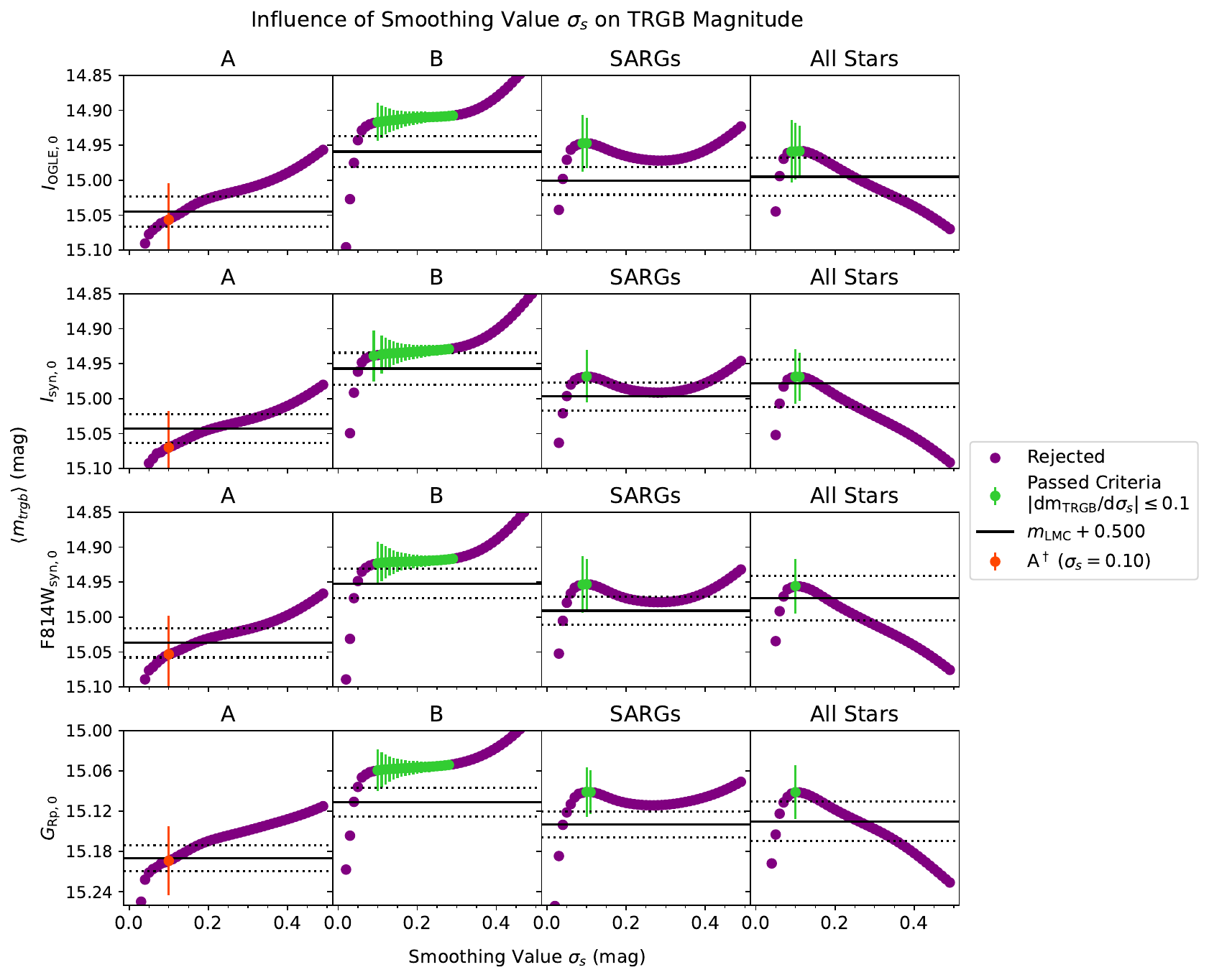}
    \caption{Mean TRGB magnitudes and standard deviations from the Monte Carlo simulations described in Section~\ref{sec:MC} over the smoothing parameter $\sigma_s$. The green points indicate the range of smoothing values insensitive to smoothing bias. For \Bseq, this range is \sigs$ \in [0.10,0.29]$, whereas it is $\sim 0.10$\,mag for the \varstars\ and \AllStars\ samples. The horizontal black lines indicate the LMC TRGB measured in A24 shifted by the difference in distance from the DEBs, \mudiff. $^\dagger:$ No stable smoothing range was found for the \Aseq\ sample in the SMC, so we adopted $\sigma_s = 0.10$\,mag.}
    \label{fig:smoothing}
\end{figure*}

\begin{table*}[ht!]
      \caption{TRGB magnitudes for the LMC and SMC for each sample and photometric band.}
      \begin{tabular}{lccccccc}
      \toprule
     &  & \multicolumn{2}{c}{$m_{\mathrm{TRGB}}$} & $\Delta \mu_{\mathrm{TRGB}}$ & \multicolumn{2}{c}{$M_{\mathrm{TRGB}}$} \\
     & & LMC & SMC & SMC-LMC & LMC & SMC \\
    Passband  & Sample  & (mag) & (mag) & (mag) & (mag) & (mag) \\
    \midrule
    $I_{\mathrm{OGLE},0}$ & A & 14.545 $\pm$ 0.013 & 15.056 $\pm$ 0.052$^\dagger$ & 0.511 $\pm$ 0.054$^\dagger$ & -3.932 $\pm$ 0.014 & -3.921 $\pm$ 0.055$^\dagger$ \\
      & B & 14.459 $\pm$ 0.014 & 14.911 $\pm$ 0.009 & 0.452 $\pm$ 0.017 & -4.018 $\pm$ 0.015 & -4.066 $\pm$ 0.018 \\
      & SARGs & 14.501 $\pm$ 0.010 & 14.947 $\pm$ 0.038 & 0.446 $\pm$ 0.040 & -3.976 $\pm$ 0.011 & -4.030 $\pm$ 0.041 \\
      & AllStars & 14.495 $\pm$ 0.021 & 14.959 $\pm$ 0.041 & 0.464 $\pm$ 0.046 & -3.982 $\pm$ 0.021 & -4.018 $\pm$ 0.044 \\
    \midrule
    $I_{\mathrm{syn},0}$ & A & 14.543 $\pm$ 0.012 & 15.070 $\pm$ 0.053$^\dagger$ & 0.527 $\pm$ 0.054$^\dagger$ & -3.934 $\pm$ 0.013 & -3.907 $\pm$ 0.055$^\dagger$ \\
      & B & 14.457 $\pm$ 0.015 & 14.933 $\pm$ 0.010 & 0.476 $\pm$ 0.018 & -4.020 $\pm$ 0.016 & -4.044 $\pm$ 0.019 \\
      & SARGs & 14.497 $\pm$ 0.011 & 14.968 $\pm$ 0.037 & 0.471 $\pm$ 0.039 & -3.980 $\pm$ 0.012 & -4.009 $\pm$ 0.041 \\
      & AllStars & 14.478 $\pm$ 0.029 & 14.969 $\pm$ 0.037 & 0.491 $\pm$ 0.047 & -3.999 $\pm$ 0.029 & -4.008 $\pm$ 0.040 \\
    \midrule
     $\mathrm{F814W}_{\mathrm{syn,0}}$ & A & 14.537 $\pm$ 0.012 & 15.053 $\pm$ 0.055$^\dagger$ & 0.516 $\pm$ 0.056$^\dagger$ & -3.940 $\pm$ 0.013 & -3.924 $\pm$ 0.057$^\dagger$ \\
      & B & 14.452 $\pm$ 0.013 & 14.920 $\pm$ 0.010 & 0.468 $\pm$ 0.016 & -4.025 $\pm$ 0.014 & -4.057 $\pm$ 0.019 \\
      & SARGs & 14.491 $\pm$ 0.010 & 14.953 $\pm$ 0.038 & 0.462 $\pm$ 0.039 & -3.986 $\pm$ 0.011 & -4.024 $\pm$ 0.041 \\
      & AllStars & 14.473 $\pm$ 0.027 & 14.956 $\pm$ 0.039 & 0.483 $\pm$ 0.047 & -4.004 $\pm$ 0.027 & -4.021 $\pm$ 0.042 \\
    \midrule
     $G_{\mathrm{RP},0}$ & A & 14.690 $\pm$ 0.009 & 15.194 $\pm$ 0.051$^\dagger$ & 0.504 $\pm$ 0.052$^\dagger$ & -3.787 $\pm$ 0.010 & -3.783 $\pm$ 0.054$^\dagger$ \\
      & B & 14.607 $\pm$ 0.013 & 15.055 $\pm$ 0.010 & 0.448 $\pm$ 0.016 & -3.870 $\pm$ 0.014 & -3.922 $\pm$ 0.019 \\
      & SARGs & 14.640 $\pm$ 0.009 & 15.091 $\pm$ 0.035 & 0.451 $\pm$ 0.036 & -3.837 $\pm$ 0.010 & -3.886 $\pm$ 0.038 \\
      & AllStars & 14.635 $\pm$ 0.024 & 15.092 $\pm$ 0.040 & 0.457 $\pm$ 0.047 & -3.842 $\pm$ 0.024 & -3.885 $\pm$ 0.043 \\
    \bottomrule
    \end{tabular}       
       \tablecomments{\mt\ is the median of the Monte Carlo TRGB magnitudes measured across the range of smoothing values that pass our criteria described in Section~\ref{sec:MC}. Uncertainties are the median standard deviation from the Monte Carlo samples. The \Bseq\ has by far the most precise TRGB measurement in the SMC. \Ad\ did not pass our smoothing-bias criteria and uses a fixed value of $\sigma_s=0.10$ mag for consistency with the LMC. $\Delta \mu_{\rm TRGB}$ is the simple difference between the apparent magnitudes per sample and passband of the SMC and the LMC. Absolute magnitude values are described in  Sect.\,\ref{sec:absmag}, and Sect.\,\ref{sec:sys_unc} provides an overview of statistical and systematic uncertainties. }
      \label{tab:AllTRGBS}
\end{table*}

 Figure~\ref{fig:smoothing} illustrates the TRGB measurements obtained as a function of $\sigma_s$ for different samples and photometric datasets; these results are listed numerically in Tab.\,\ref{tab:AllTRGBS} alongside the LMC results presented in A24. The smoothing insensitive range of $\sigma_s$ over which we determine the final values of \mtrgb\ is highlighted using green circles with errorbars that correspond to \sigmc. As in A24, we find that the shape of the observed LF has a significant impact on smoothing bias. Additionally, the same hierarchy of magnitudes is found in the SMC as was reported in the LMC by A24: the \Bseq\ sample consistently yields the brightest \mtrgb, followed by \sargs\ (and \AllStars), and the \Aseq\ sample always yields the faintest \mtrgb. \sargs\ and \AllStars\ results are fully consistent with each other, as expected from the fact that all red giants near the TRGB are also \sargs\ (Fig.\,\ref{fig:CMD}). As in A24, we find that the \AllStars\ LFs is rather sensitive to smoothing bias and that only a small range of $\sigma_s$ values yields consistent TRGB magnitudes.

The \Bseq\ sample yields the best TRGB measurement over the largest range of $\sigma_s$ in the SMC, whereas the \sargs\ sample had provided the best overall measurement in the LMC (A24). Specifically, we measured \mio\ $=14.911 \pm 0.009$\,mag for the \Bseq, which is slightly more precise than the \Bseq\ in the LMC and subject to a systematic uncertainty of $0.008$\,mag (Tab.\,\ref{tab:errors}).  The gain in precision for the \Bseq\ TRGB measurement is driven by the higher degree of smoothing, which by construction boosts the Sobel filter signal \citep{Hatt17}. We use the \Bseq\ as our baseline for further inspections of systematics such as metallicity and reddening effects.

Contrary to the LMC, the SMC's \Aseq\ sample is particularly sensitive to smoothing bias and there is indeed no range of $\sigma_s$ where the dependence of \mtrgb\ on $\sigma_s$ is flat. As a result, we consider the \Aseq\ measurement less reliable and for comparison only using a fixed value of $\sigma_s=0.10$\,mag, which corresponds to the typical smoothing value for the LMC \Aseq\ (A24) and is not far from the typical combined photometric uncertainty. This yields \mio$=15.056 \pm 0.052$\,mag, with an additional systematic uncertainty of $0.015$\,mag (Tab.\,\ref{tab:errors}). 

We find similar values of \mtrgb\ for the \sargs\ and \AllStars\ samples of \mio$=14.947\pm0.038$\,mag and \mio$=14.959\pm0.041$\,mag, respectively. Similar to the \Aseq, we find a very restricted 
$\sigma_s$ range for the \sargs\ and \AllStars\ samples. This is likely the case because \Aseq\ stars are much more numerous than \Bseq\ stars (cf. Tab.\,\ref{tab:color}) and hence dominate the LF starting at approximately $0.10$\,mag below the \Bseq\ \mtrgb.

\begin{table}[]
    \caption{Difference in TRGB brightness between samples}
    \hspace{-1cm}
    \begin{tabular}{lccc}
    \toprule
        Band & LMC & SMC & diff \\
         \midrule
         & \multicolumn{2}{c}{$m_{Bseq}-m_{\mathrm{SARGs}}$} &   \\
        \midrule
        $I_{\mathrm{OGLE},0}$ & -0.042 $\pm$ 0.017 & -0.036 $\pm$ 0.039 & 0.1$\sigma$ \\
        $I_{\mathrm{syn},0}$ & -0.040 $\pm$ 0.019 & -0.035 $\pm$ 0.039 & 0.1$\sigma$ \\
        $\mathrm{F814W}_{\mathrm{syn,0}}$ & -0.039 $\pm$ 0.016 & -0.033 $\pm$ 0.039 & 0.1$\sigma$ \\
        $G_{\mathrm{Rp},0}$ & -0.033 $\pm$ 0.016 & -0.036 $\pm$ 0.036 & 0.1$\sigma$ \\
        \midrule
         & \multicolumn{2}{c}{$m_{\rm Bseq}-m_{\rm Aseq}$} & \\
         \midrule
        $I_{\mathrm{OGLE},0}$ & -0.086 $\pm$ 0.019 & -0.146 $\pm$ 0.053$^\dagger$ & 1.1$\sigma$$^\dagger$ \\
        $I_{\mathrm{syn},0}$ & -0.086 $\pm$ 0.019 & -0.137 $\pm$ 0.054$^\dagger$ & 0.9$\sigma$$^\dagger$ \\
        $\mathrm{F814W}_{\mathrm{syn,0}}$ & -0.085 $\pm$ 0.018 & -0.133 $\pm$ 0.056$^\dagger$ & 0.8$\sigma$$^\dagger$ \\
        $G_{\mathrm{Rp},0}$ & -0.083 $\pm$ 0.016 & -0.139 $\pm$ 0.052$^\dagger$ & 1.0$\sigma$$^\dagger$ \\
        \bottomrule
    \end{tabular}
    \tablecomments{\textit{Top:} Comparison of $m_{\rm Bseq}-m_{\mathrm{SARGs}}$ (mag) for the LMC and SMC alongside the agreement observed between the differences in both galaxies.  The difference is consistent in each of the photometric bands considered and in both galaxies. \textit{Bottom:} Comparison of $m_{\rm Bseq}-m_{\rm Aseq}$ for the LMC and SMC. $^\dagger$ identifies that the SMC \Aseq\ does not satisfy our smoothing bias criteria.}
    \label{tab:BvsSARGsAndA}
\end{table}

Table\,\ref{tab:BvsSARGsAndA} compares apparent magnitude differences between the samples in both galaxies and shows that the \Bseq\ sample yields a $\sim 0.04$\,mag brighter \mtrgb\ magnitude than the \sargs\ in both the SMC and the LMC. The \Bseq\ is furthermore brighter than the \Aseq\ by $0.15 \pm 0.05$\,mag in the SMC and $0.09 \pm 0.02$\,mag in the LMC. This brightness difference is readily apparent also in the LFs in Fig.~\ref{fig:PL} and established independently of distance, with similar differences in every photometric band considered ($\lesssim 0.01$\,mag variation). Therefore we conjecture that this brightness difference is of astrophysical origin and can be found in other galaxies as well.

Interestingly, we find that \mio\ is $\sim0.01-0.02$ mag brighter than \mic\ in the SMC. Conversely, in the LMC \mic\ is slightly brighter than \mio, although both agree to within a few mmag. The difference is more pronounced for \AllStars. Considering that the LMC stars near the Tip are redder than the SMC, the difference between the SMC and LMC samples is consistent with the color-dependent photometric system differences seen in Figure~\ref{fig:GaiaVsOGLEIband}. Moreover, the difference between \mio\ and \mic\ in the SMC matches the star-to-star comparison presented in Section~\ref{sec:Ioglevgaia} above, which yields an average difference of $0.022$\,mag.

\subsection{Spatial analysis shows radial metallicity gradient\label{sec:spatial}}\label{sec:metallicity_effects}
\begin{figure*}
    \centering
    \includegraphics[width=1\textwidth]{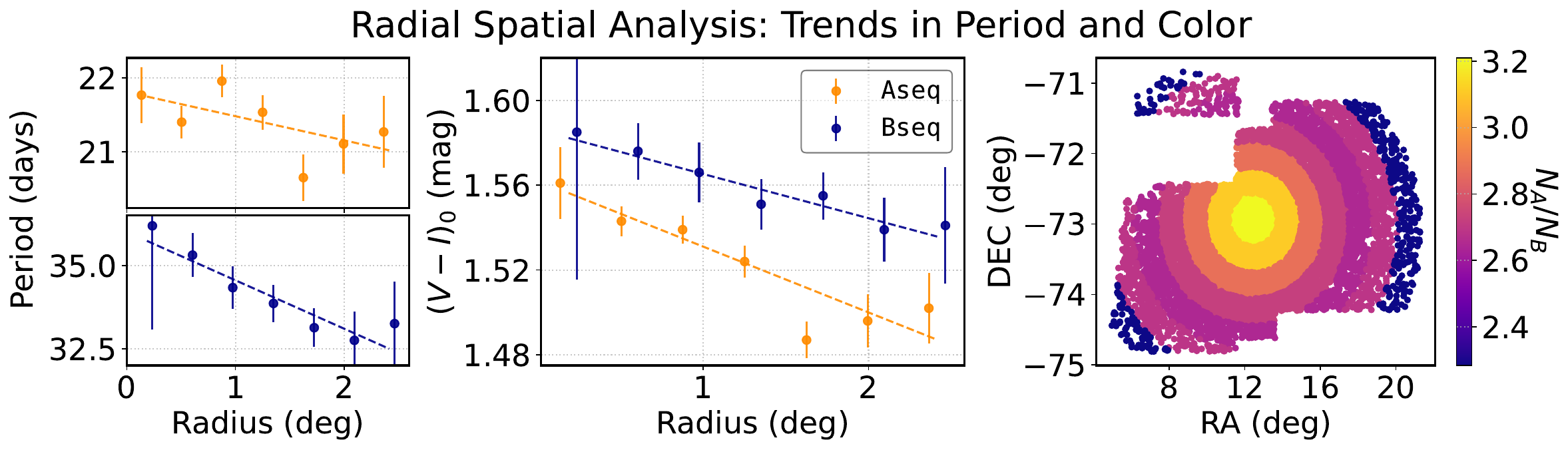}
    \caption{Properties of the \Aseq\ and \Bseq\ samples (magnitude range 14.9 to 15.1\,mag) assessed within rings of increasing radius centered on the SMC core.  \textit{Left:} Median period versus the mid-radius of each ring for stars in \Aseq\ and \Bseq. Median periods get longer moving towards the the core. Error bars show the standard error on median. \textit{Center:} Median de-reddened $(V-I)_0$ color versus ring mid-radius for all sequences, revealing that stars get redder towards the core.  \textit{Right:} Spatial distribution of the rings in the SMC colored by the ratio of \Aseq\ to \Bseq\ stars showing more \Aseq\ stars in the core. The center point was taken to be RA: 12.44$^\circ$ and DEC: -72.94$^\circ$ \citep{Graczyk2020}. Fit parameters are reported in Table~\ref{tab:radial_fits}.} 
    \label{fig:radialAB}
\end{figure*}

\begin{figure*}
    \centering
    \includegraphics[width=1\textwidth]{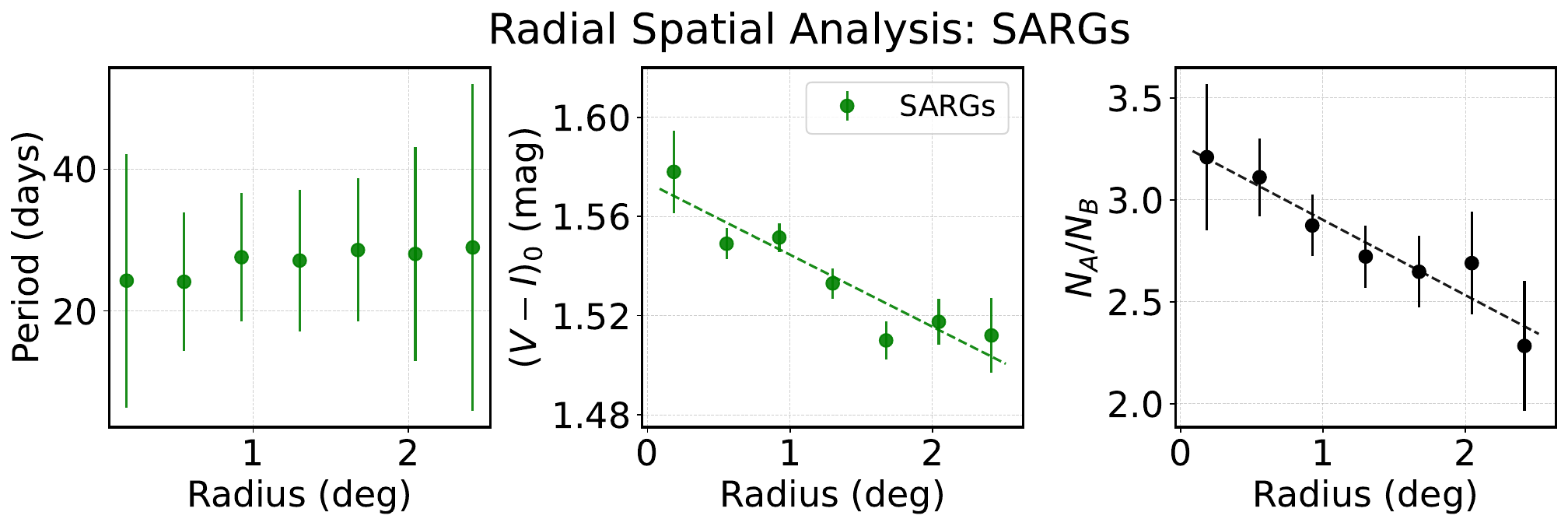}
    \caption{Radial spatial analysis based on all \sargs\ in the SMC, that is, without distinguishing sequences. \textit{Left:} Median period versus radius is nearly constant for \sargs. \textit{Center:} Median color versus ring mid-radius shows that \sargs\ become redder in the core. \textit{Right:} Number of \Aseq\ stars relative to \Bseq\ stars. Isolating A and B provided us the ability to discern the period trend in Fig.\,\ref{fig:radialAB}, which disappears when considering the  median periods of all \varstars\ because their relative number is not constant. Error bars represent the standard error on median. Fit parameters are reported in Table~\ref{tab:radial_fits}.
    \label{fig:radialSARGs}}
\end{figure*}

The SMC's metallicity gradient has been traced using RGB stars \citep{Povick2023SMC, Li2024} and classical Cepheids \citep[their Fig.\,5]{Breuval2022}. Here, we show that \sargs\ also trace the SMC's metallicity gradient. We grouped stars into spatial rings based on their radial separation from the SMC center (12.44$^\circ$, -72.94$^\circ$) \citep{Graczyk2020}. The annular regions were defined at 7 equidistant intervals out to a maximum radius of 2.6 degrees from the core. The mid-radius, i.e., the region half-way between inner and outer boundary, is used to quantify the radial trends. 

We computed the median period $P_1$ and median color $(V-I)_0$ for each ring for stars with $I-$band magnitudes between $14.9$ and $15.1$\,mag for all three samples involving \sargs. While we also attempted to measure the TRGB magnitude of each ring, noise due to insufficient number of stars complicates detecting a significant trend in \mtrgb. 

As Figure~\ref{fig:radialAB} shows, the median period and $(V-I)_0$ color of both the \Aseq\ and \Bseq\ samples increase towards the SMC's core. This suggests that RGs in the SMC's core have higher metallicity, which causes them to appear redder and exhibit larger radii due to line blanketing. Given a nearly constant mass, larger radii result in longer periods. Furthermore, the ratio of $N_A/N_B$ increases towards the center, implying an age gradient since \Aseq\ stars are younger than \Bseq\ stars (cf. Figure~\ref{fig:Age} and Table~\ref{tab:color}). Both trends suggest a younger and more metal-rich core, likely a consequence of the SMC's cumulative star formation history and in agreement with the negative metallicity gradient reported using RGs of different ages in \citet[their Fig.\,10]{Povick2023SMC}.  Hence, the variability of RG near the TRGB provides useful information for inferring the properties of the RG  populations. 

Figure~\ref{fig:radialSARGs} shows the analogous radial analysis based on the \sargs\ sample, which does not distinguish between P-L sequences and is mostly composed of \Aseq\ and \Bseq\ stars. The ratio of the number of stars on the two sequences is seen to increase towards the center, as is the average color. However, there is no radial trend with average period because the relative increase in shorter-period \Aseq\ stars towards the center compensates the trends with period that are present on both sequences. These trends can only be recovered by selecting samples according to their period-luminosity sequences. Thus, while color traces metallicity trends regardless of the RG sample, tracing ages according to frequencies requires considering the dominant periods of the RGs. Table~\ref{tab:radial_fits} lists the fitted trends from Figures~\ref{fig:radialAB} and \ref{fig:radialSARGs}.

\begin{table}
    \caption{Fitted spatial trends in period, color, and $N_A/N_B$}
    \hspace{-2cm}
    \begin{tabular}{lcccccc}
    \toprule
         & \multicolumn{2}{c}{$P_1$ } & \multicolumn{2}{c}{$(V-I)_0$} & \multicolumn{2}{c}{$N_A/N_B$} \\
        & \multicolumn{2}{c}{(days)} & \multicolumn{2}{c}{(mag)} \\
        Sample & m & c & m & c & m & c \\
        \midrule
        \Aseq & -0.33 & 21.81 & -0.031 & 1.56 &  &  \\
        \Bseq & -1.45 & 36.0 & -0.021 & 1.59 &  &  \\
        \sargs\ &  &  & -0.03 & 1.57 & -0.37 & 3.27 \\
\bottomrule
    \end{tabular}
    \label{tab:radial_fits}
    \tablecomments{Linear fit parameters as shown in Figures~\ref{fig:radialAB} and \ref{fig:radialSARGs}, with slope $m$ and intercept $c$.}
\end{table}

\subsection{A period-color relation at the TRGB}\label{sec:periodcolor}
\begin{figure*}[ht!]
    \includegraphics[width=1\textwidth]{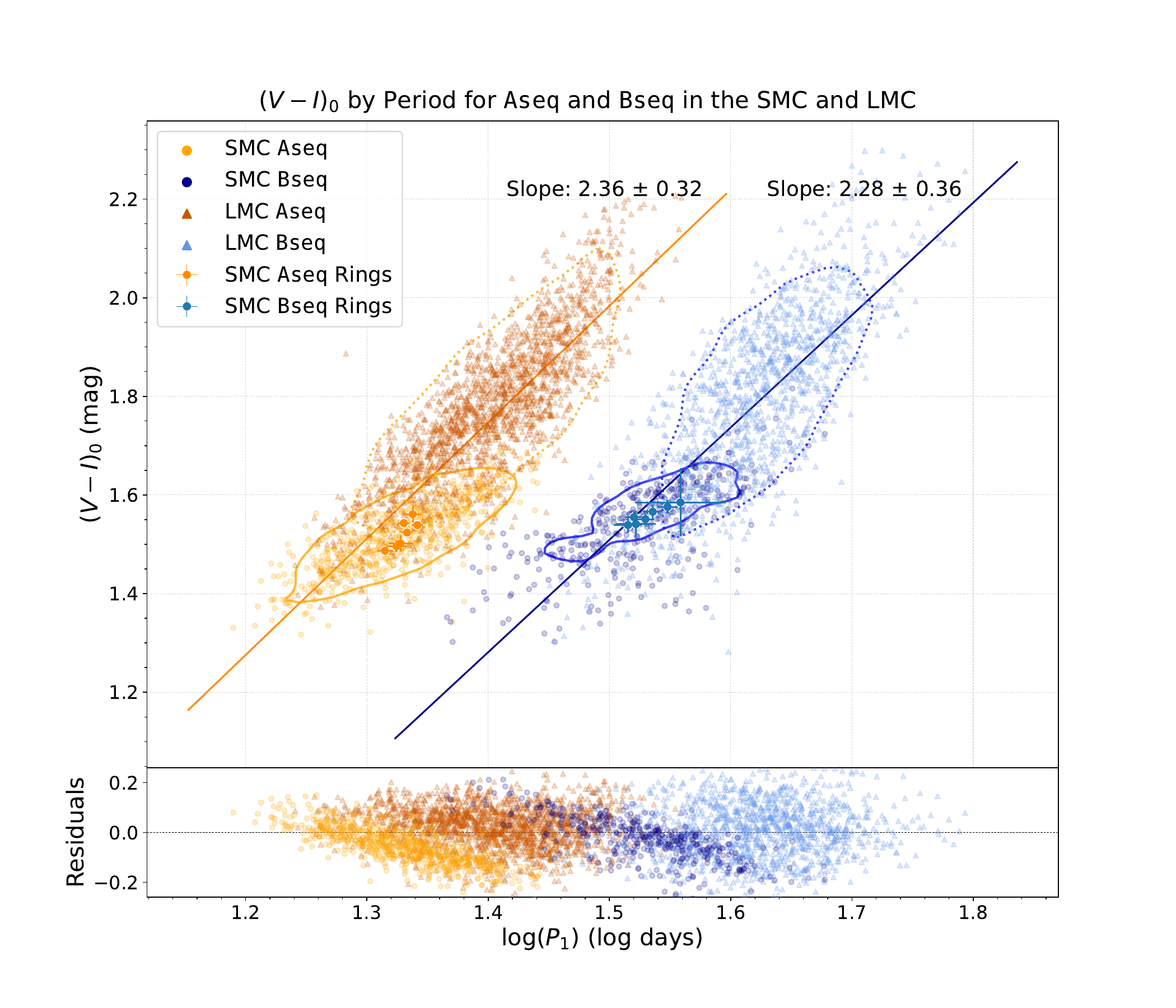} 
    \caption{Period-color relation for \Aseq\ and \Bseq\ \sargs\ in the LMC and SMC within 0.1 mag of \mtrgb. Fits based on both galaxies reveal a statistically significant period-color relation. The residual trends reveal differences in slope between the galaxies that could be due to differing ages and metallicities. The median periods and colors of the SMC spatial rings and the LMC are shown as 2D errorbars for reference only.}
    \label{fig:PeriodcolorRelationship}
\end{figure*}

Stars near the RGB tip are intrinsically redder in the LMC than in the SMC due to the latter's lower metallicity, cf. Tab.~\ref{tab:color}. Additionally, the periods of both \Bseq\ and \Aseq\ stars near the TRGB are shorter in the SMC than the LMC, which implies smaller radii for equal mass, since $P\propto1/\sqrt{\rho}$ \citep{Ritter1879,Rosseland1943}, similar to the radial analysis shown above. The consequence of both (color and period) trends reverting to the same physical origin (metallicity) is the existence of a metallicity-sensitive period-color relation for small amplitude red giants near the RGB tip. 

Figure~\ref{fig:PeriodcolorRelationship} illustrates this period-color relation for the TRGB using stars from the \Aseq\ and \Bseq\ samples that are within $0.1$\,mag of the respective \mtrgb. For the SMC, we also show average values reported for the different annular regions considered in Section \ref{sec:spatial} as larger errorbars. Overall, we see two rather clean sequences spanning approximately $0.4$ dex in $\log{P}$ and up to $1.0$\,mag in $(V-I)_0$ color. While there is overlap among the stars from both galaxies, the SMC stars tend to populate the shorter-period, bluer parts of the relations, and the LMC stars the redder, longer-period parts. Remarkably, stars from both galaxies appear to mostly follow a single relation per SARG sequence. Combined fits to the SMC and LMC stars yield $(V-I)_0=(2.36\pm0.32)(\log{P_1}-1.4)+(1.75\pm0.02)$ mag for the \Aseq\ and $(V-I)_0=(2.28\pm0.36)(\log{P_1}-1.6)+(1.74\pm0.02)$ mag for the \Bseq. 

The TRGB's period-color relations provide a potentially useful tool for addressing interstellar reddening, for example for calibrating \Mt\ based on \gaia\ parallaxes of Milky Way RGs whose SARG-like variability will be classified in increasing detail in future \gaia\ data releases by the long-period variables specific object studies \citep{2023A&A...674A..15L,2023A&A...674A..13E}.
Additionally, Section\,\ref{sec:absmag} below considers the use of these period-color relations to standardize TRGB magnitudes for metallicity differences in the absence of spectroscopic information.

\subsection{Distance modulus differences measured with and without a metallicity correction}\label{sec:mudiff}

We computed the difference in distance modulus, \mudiff, using the values of \mtrgb\ determined for the \Bseq\ and \sargs\ samples in Section\,\ref{sec:mtrgb} and listed in Tab.\,\ref{tab:AllTRGBS}. If the absolute magnitude of the TRGB were identical in both galaxies, then one would expect to find a value of \mudiff $\approx 0.500$\,mag as measured using DEBs \citep{Graczyk2020}. However, we found that \mudiff\ tends to be $1-2\sigma$ smaller than the reference number, irrespective of the sample or photometric bands considered. We therefore decided to investigate whether metallicity or dust effects could improve the agreement. The results are listed in Table\,\ref{tab:corrected_dmu}.

\begin{table*}[]
    \caption{Comparison of corrected distance moduli for the LMC and SMC using $I_{\mathrm{OGLE},0}$.}
    \begin{tabular}{lccc}
    \toprule
              &                      & \mudiffogle & \mudiffogle \\
    Sample & $R_{V,\mathrm{SMC}}$ &  & $-0.217\Delta_{SMC-LMC}(V-I)_0$ \\
    \midrule
     \Aseq & 2.7 & 0.511 ± 0.054 (0.2$\sigma$) & 0.569 ± 0.054 (1.2$\sigma$) \\
     & 3.3 & 0.489 ± 0.060 (0.2$\sigma$) & 0.547 ± 0.060 (0.8$\sigma$) \\
    \midrule
    \Bseq & 2.7 & 0.452 ± 0.017 (2.0$\sigma$) & 0.513 ± 0.018 (0.5$\sigma$) \\
     & 3.3 & 0.443 ± 0.018 (2.3$\sigma$) & 0.504 ± 0.019 (0.2$\sigma$) \\
    \midrule
    \sargs & 2.7 & 0.446 ± 0.040 (1.2$\sigma$) & 0.501 ± 0.040 (0.03$\sigma$) \\
     & 3.3 & 0.443 ± 0.038 (1.4$\sigma$) & 0.498 ± 0.039 (0.04$\sigma$) \\
    \midrule
    \AllStars & 2.7 & 0.464 ± 0.046 (0.7$\sigma$) & 0.520 ± 0.046 (0.4$\sigma$) \\
     & 3.3 & 0.457 ± 0.042 (0.9$\sigma$) & 0.513 ± 0.043 (0.3$\sigma$) \\
    \bottomrule
\end{tabular}
    \tablecomments{$\Delta \mu$ is computed as the straight difference in apparent magnitudes in the OGLE-III $I-$band. The comparison is most informative for the \Bseq\ thanks to its higher precision. The other differences are used to assess reddening systematics in Tab.\,\ref{tab:errors}. In parentheses, we report the agreement in $\sigma$ with $\Delta \mu_{\mathrm{SMC-LMC}} = 0.500 \pm 0.017$ from \cite{Graczyk2020}. The last column presents the apparent magnitude difference between both galaxies, corrected for metallicity using color following \citet{Rizzi2007}. For the precise \Bseq, the metallicity corrected $\Delta \mu$ improves from $2.0-2.3\sigma$ to $0.5-0.2\sigma$.}
    \label{tab:corrected_dmu}
\end{table*}

The color-based metallicity corrections by \citet{Rizzi2007} improve the agreement between the expected and measured \mudiff\ values. Indeed, color-corrected distance differences, $\Delta \mu_{\mathrm{SMC-LMC,corrected}} = \Delta \mu_{\mathrm{SMC-LMC}} - 0.217((V-I)_{0,\mathrm{SMC}}-(V-I)_{0,\mathrm{LMC}})$, agree nearly perfectly with the expected value based on DEBs: the \Bseq\ SMC-LMC distance modulus, \mudiffogle\, agrees with DEBs to within 0.5$\sigma$ after metallicity corrections, and \sargs\ agree with DEBs to within 0.1$\sigma$.

For comparison, changing the reddening law for the SMC has a much smaller effect on \mudiff. Using $R_{V,\mathrm{SMC}} = 3.3$ instead of our default \citep[$R_V=2.7$][]{Gordon2003} increases the difference between the \mudiff\ values only very slightly (by less than $0.01$\,mag) because of the low reddening (typical E(V-I)=0.05\,mag) of the SMC.  As an aside, the scatter in the reddening-free Wesenheit-magnitudes for the \Bseq\ increases from 0.176 to 0.182 when assuming $R_{\mathrm{V,SMC}}=3.3$. We therefore kept $R_V=2.7$ and adopted half the range of the reddening-law related differences as part of the systematic uncertainties stated in Tab.\,\ref{tab:errors}.

\subsection{Absolute TRGB magnitudes and standardization \label{sec:absmag}}

\begin{figure*}
    \centering
    \includegraphics[width=1\textwidth]{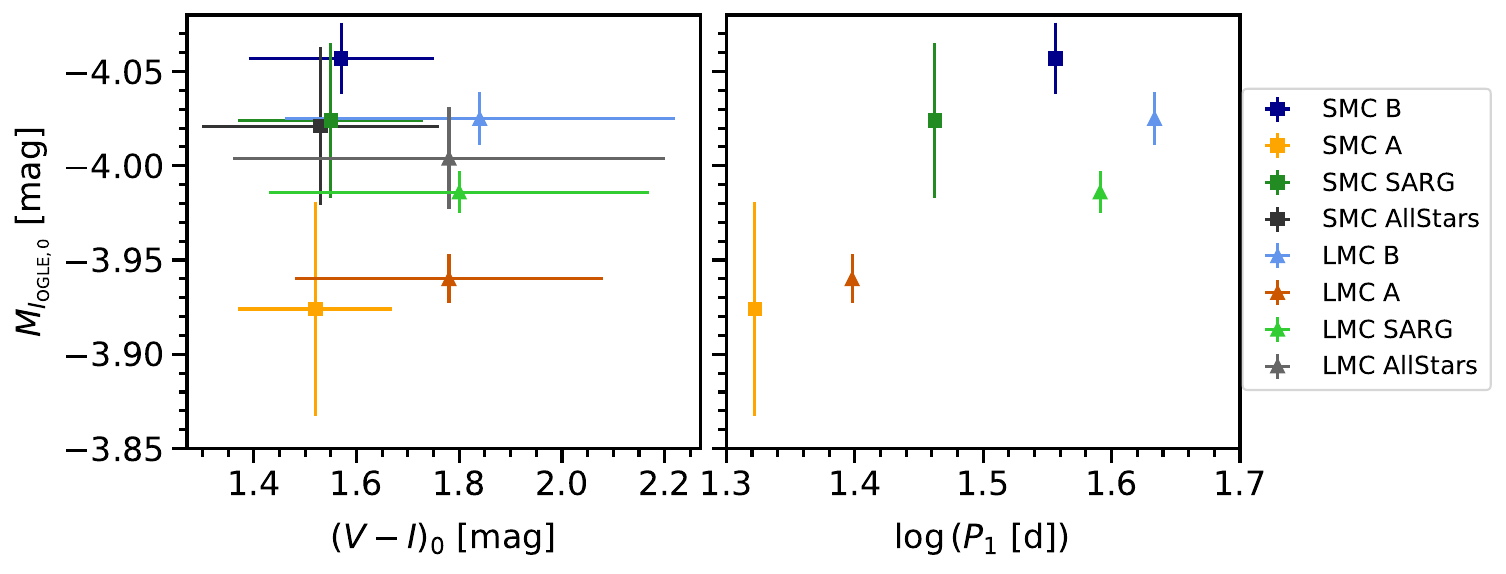}
    \caption{The LMC and SMC show the same hierarchy of \Mt\ among the samples considered: \Bseq\ is brightest, \sargs\ and \AllStars\ yield similar results (100\% of stars at the TRGB are SARGs), and the \Aseq\ yields the faintest TRGB magnitude. Vertical uncertainties shown do not include the systematic uncertainty of the LMC and SMC distances.
 {\it Left:} \ogle\ $I-$band absolute magnitude as a function of color, with SMC stars being bluer than LMC stars. {\it Right:} Periods are longer in the LMC compared to the SMC for each of the samples. The absolute magnitudes in both Clouds increase with the mean period of the samples, and the SMC tends to be slightly brighter, albeit not significantly so.\label{fig:MabsDiversity}}
\end{figure*}

\begin{figure}[ht!]
    \includegraphics[width=0.5\textwidth]{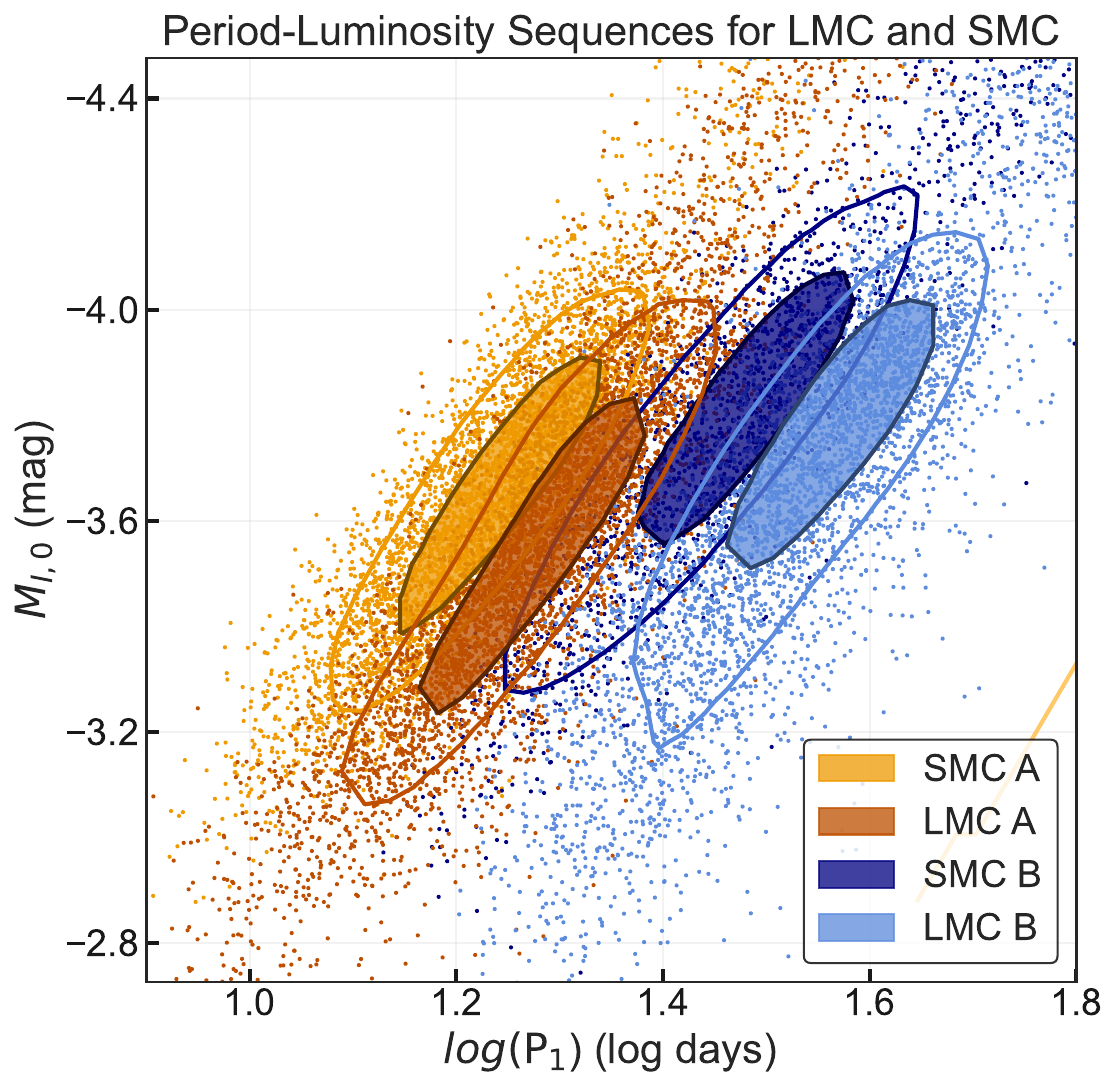}
    \caption{Period-luminosity sequences for both the LMC and SMC in absolute magnitude after correcting for extinction $M_{I,0}$. The contours show a noticeable shift of both A and B sequences towards both shorter period and higher luminosity in the more metal-poor SMC compared to the LMC.}
    \label{fig:PL_SMCLMCShift}
\end{figure}

Adopting the geometric distance to the SMC determined using DEBs \citep{Graczyk2020}, we translate the measured apparent magnitudes, \mt, into absolute magnitudes. The most accurate of these measurements is obtained for the \Bseq, with \Mio$\, =-4.066 \pm 0.018 \mathrm{(stat.)} \pm 0.029 \mathrm{(syst.)}$ at a mean color of $(V-I)_0 = 1.57$ mag. This is $0.041$\,mag brighter then the \Bseq\ TRGB calibration in the LMC, where A24 reported $-4.025 \pm 0.014 \mathrm{(stat.)} \pm 0.33 \mathrm{(syst.)}$\,mag at slightly redder color. Figure\,\ref{fig:MabsDiversity} compares the results for the different samples obtained in the LMC and SMC as a function of color and average $\log{P_1}$ and illustrates the hierarchy of the tip magnitudes belonging to the different samples. Figure\,\ref{fig:PL_SMCLMCShift} shows the period-absolute $I-$band magnitude relations for the \Aseq\ and \Bseq\ stars in both galaxies. The SMC's shorter periods and brighter RGB tips are readily apparent even just from the contours of the stars on these sequences.

\begin{figure*}[ht!]
    \centering
    \includegraphics[width=0.75\textwidth]{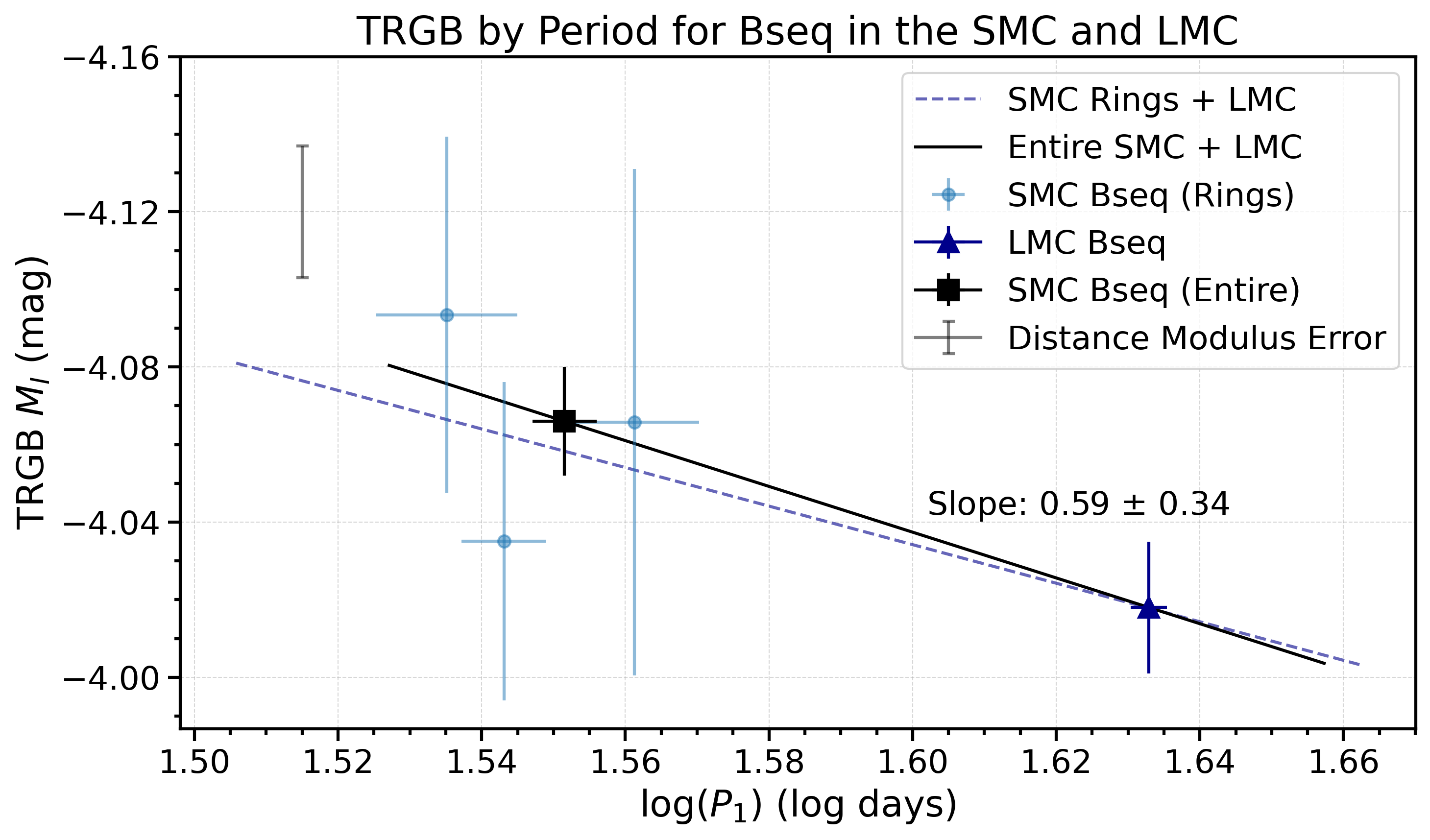} 
    \caption{Absolute TRGB magnitude ($M_I$) as a function of the median logarithmic period ($\log(P_1)$) for the \Bseq\ stars in the LMC, SMC, and SMC spatial rings. The TRGB depends on period by $M_I = (0.59 \pm 0.34 \frac{\mathrm{mag}}{\mathrm{log days}}) (\log(P_1) - 1.63) + (-4.018 \pm 0.015 \pm 0.027)$ mag when using the the LMC \Mio\ (Tab.~\ref{tab:AllTRGBS}) and period (Tab.~\ref{tab:color}). This in principle allows to create a period-based metallicity correction, in analogy with known color-based metallicity calibrations. Unfortunately, however, the metallicity difference between the SMC and LMC is small, preventing an accurate calibration of the period-metallicity relation at the TRGB. 
    The slope, $0.59 \pm 0.34 \frac{\mathrm{mag}}{\mathrm{log days}}$, is calculated using OGLE-III $I-$band as $(m_{TRGB,B,SMC} - m_{TRGB,B,LMC} - \Delta \mu_{\mathrm{SMC-LMC, Graczyk}})/(\log{P_1}_{SMC} - \log{P_1}_{LMC})$. The $\Delta \mu_{\mathrm{SMC-LMC}}$ error from \cite{Graczyk2020} is plotted for reference in the top left corner.}
    \label{fig:TRGBvPeriod}
\end{figure*}

Calculating the difference in $M_I$ vs $\log{P}$ for the \Bseq\ yields a slope of $0.59 \pm 0.34$\,mag/dex\footnote{`dex' in this paragraph refers to a decade in logarithmic period}, which at least in principle could be used to correct metallicity differences in $M_I$ using variability periods instead of color, see Figure\,\ref{fig:TRGBvPeriod}. Unfortunately, the uncertainty on this slope is rather large due to the small metallicity difference between the LMC and SMC and due to the significant dispersion of the SARG period-luminosity relations. Nevertheless, we combined the period-$M_I$ relation with the period-color relation of Sect.\,\ref{sec:periodcolor} ($\mathrm{(V-I)_0} \propto (2.28 \pm 0.36) \mathrm{mag/dex}$) to determine the implied color-based metallicity correction of \Mio $\propto (0.26 \pm 0.16) \cdot \mathrm{(V-I)_0}$. This is in good agreement with \Mi\ $\propto 0.217(V-I)_0$ from \citet{Rizzi2007} and \Mi\ $\propto (0.2 \pm 0.05)(V-I)_0$ from \citet{Madore2009}, although it agrees somewhat less with \Mio\ $\propto (0.053\pm0.019)(V-I)_0$ reported by \citet{Hoyt2023}. Greater leverage on metallicity is needed to improve this period-based metallicity correction of TRGB magnitudes, which could have the benefit of being independent of color, and hence, reddening. The combined, period-dependent absolute magnitude calibration of the \Bseq\ TRGB across both galaxies using periods to correct for metallicity differences thus becomes:
\begin{equation}\label{eq:periodcalibration}
M_{\mathrm{I,OGLE}} = M_\mathrm{I, OGLE, 0} + (0.59 \pm 0.34) \mathrm{mag} \cdot \log{(P/P_0)} \ , 
\end{equation}
with $P_0 =42.7$\,d and $M_\mathrm{I, OGLE, 0} = -4.018 \pm 0.015 \mathrm{(stat.)} \pm 0.027 \mathrm{(syst.)}$\,mag based on the LMC calibration, cf. Tab.\,\ref{tab:AllTRGBS} and A24. 

We note that our \AllStars\ TRGB magnitude in the SMC, \Mio $ = -4.018 \pm 0.044 \mathrm{(stat.)} \pm 0.029 \mathrm{(syst.)}$\,mag, is slightly dimmer than the value of $-4.050 \pm 0.021 \mathrm{(stat.)} \pm 0.038 \mathrm{(syst.)}$ mag reported by \citet{Hoyt2023} based on the same dataset. While the difference between the two apparent magnitudes ($0.032$\,mag) agrees within the reported uncertainties, we briefly mention some of the methodological differences in Sec.~\ref{sec:otherstudies} and defer to Appendix C in A24 for further detail. Here, we simply point out that the two numbers should not be compared without considering the substantial systematic differences between the present work and \citet{Hoyt2023}.

Although the \Bseq\ provides the most accurate and luminous calibration in the SMC, we stress that the use of this calibration for measuring distances is a) limited to galaxies where the \Bseq\ sample can be selected based on variability periods (such as in the present comparison between LMC and SMC) and b) subject to a metallicity effect that should be taken into account. We stress that using the \Bseq\ calibration to measure the distance to an \AllStars-like sample of RGs in another galaxy would overestimate the distance modulus by approximately $0.04$\,mag ($2\%$ in distance), cf. Tab.\,\ref{tab:AllTRGBS}. The \sargs-based calibration of \Mih$=-4.024 \pm 0.041 \mathrm{(stat.)} \pm 0.029 \mathrm{(syst.)}$\,mag should thus be considered the most generally applicable one for distance determinations. However, the sensitivity of the absolute magnitude based on the other three samples indicates that the relative contributions of younger and older RGs on the different LPV PL-sequences can influence the result in ways that are difficult to quantify without detailed variability information.
Given that the slope of the TRGB feature depends both on wavelength and population age \citep{Valenti2004,Madore2023}, such issues will be particularly relevant for NIR-IR TRGB distances measured using the JWST \citep[e.g.,][]{Newman2024a,Newman2024b}.

\section{Discussion}\label{sec:discussion}

\subsection{Comparison to Other Studies}\label{sec:otherstudies}

Calibrations of the TRGB in the Small Magellanic Cloud based on the geometric distance by \citet{Graczyk2020} have been presented in the recent literature by \citet{Freedman2020} and \citet[henceforth: H23]{Hoyt2023}, and previously also by \citet{gorski2016} and \citet{Yuan2019}. In particular, H23 employed the same OGLE-IV reddening maps \citep{Skowron2021ApJS} and OGLE-III photometry considered by us. However, there are several significant methodological differences between our work and H23 as explained in Appendix C of A24. The most significant differences in H23 relative to this work include the use of weighted Sobel filter responses, non-specified smoothing parameter, application of color cuts, and differences in the treatment of reddening, among others. We further note that the most direct comparison between H23 and the present work should consider our results for the \AllStars\ sample, after correcting methodological differences. Suffice it here to mention that the spatial selection applied in H23 affected the reported \mtrgb\ by $0.004$\,mag (Fig. 2 in H23) and does not follow the age trend detected by variability, cf. Figs.\,\ref{fig:radialAB} and \ref{fig:radialSARGs}. Specifically, differences in sample and methodology do not allow a direct comparison of the apparent magnitude of $m_{I,0}^{\mathrm{H23}} = 14.927 \pm 0.023$\,mag reported in H23 with our \Bseq\ calibration of $m_{I,0}=14.911 \pm 0.009 \mathrm{(stat.)} \pm 0.008 \mathrm{(syst.)}$\,mag, even if the two numbers are similar. We note that the larger uncertainty of our \AllStars\ sample result ($14.959 \pm 0.041 \mathrm{(stat.)} \pm 0.008 \mathrm{(syst.)}$\,mag) originates mainly from the significant sensitivity of \mtrgb\ to \sigs\ and the treatment of reddening uncertainties, which do not average as $\sqrt{N}$ as assumed in H23 (cf. A24).

Very recently, \citet{Bellazzini2024} presented TRGB measurements in the LMC and SMC based on {\it Gaia} DR3 synthetic photometry, including the bands labeled here as $I_{\mathrm{syn}}$ (their JKC $I$) and $\mathrm{F814W_{syn}}$ (their ACS-WFC F814W). Small differences can be expected notably due to the exclusion of the inner regions of the LMC and SMC by \citet[cf. their Fig.\,1]{Bellazzini2024}. Nevertheless, despite using rather different regions of the sky to measure \mt, and despite several differences in methodology, our results for the \AllStars\ sample agree to better than $1\sigma$ with their reported apparent magnitudes. Specifically for the SMC, we find $14.969\pm0.037$\,mag vs. their $14.994 \pm 0.015$\,mag in $I_{\mathrm{syn}}$ and $14.956 \pm 0.039$\,mag vs. their $14.981 \pm 0.014$\,mag in $\mathrm{F814W_{syn}}$. The comparison for the LMC is similar. 

Since Sobel filter response weighting introduces a tip-contrast relation \citep[A24]{Wu2022}, it is crucial to standardize TRGB magnitudes measured using weighted Sobel filters. In the case of the LMC and SMC, the overall tip contrast is very similar ($N_+/N_- = 3.0$ and 3.4 respectively with a normal Sobel filter, see A24 for definitions), so that the related correction will not be significant when considering $\Delta \mu$ between the two galaxies. However, this situation can be very different for other galaxies, or even among different fields of the same galaxy \citep{Scolnic2023}, when TRGB contrasts differ more significantly.  

\subsection{Relevance for extragalactic distances and the Hubble constant\label{sec:H0}}

We underline the importance of using consistently measured \mtrgb\ values when measuring distances, e.g., with the goal of determining the Hubble constant, $H_0$. The typical desired populations for determining $H_0$ are old, metal-poor RGs \citep[e.g.,][]{Freedman2020}. Field-to-field variations of \mtrgb\ within the same galaxy \citep{Wu2022,Scolnic2023} underline the need to ensure consistency of algorithmic aspects of measuring \mtrgb\ as well as the underlying stellar population \citep[cf. also][and references therein]{Beaton2018,LiH0book2024}. As this work and A24 show, the diversity of RG populations within a given galaxy affects TRGB measurements at the level of $0.04$\,mag. Larger effects related to metallicity differences must be standardized when comparing the TRGB in different galaxies \citep{Rizzi2007}, and we show that variability periods are very sensitive to metallicity differences, yet reddening independent. While the impact of age differences cannot yet be standardized, there is hope that variability amplitudes could help if the age-amplitude reported here can be more securely established.

In principle, the Magellanic Clouds offer crucial opportunities for the absolute calibration of the TRGB as a standard candle thanks to the availability of geometric distances. However, the intermediate-age and intermediate-metallicity red giant populations of the Magellanic Clouds are close to the limits of the parameter range (age $> 4-5$\,Gyr, [M/H]$\lesssim -0.7$) considered suitable for measuring precise distances to older,  metal-poorer red giant populations targeted at greater distances according to stellar models \citep{Barker2004,Salaris2005,Cassisi2013,Serenelli2017}. Furthermore, the Magellanic Clouds are observed under different conditions than more distant galaxies, using different (typically: ground-based) telescopes and photometric systems, and using different photometric algorithms. The impact of population differences between the Magellanic RG populations and those residing in the outer halos of more distant SN-host galaxies requires further assessment.

The TRGB calibrations reported here seek to limit biases due to methodological choices. In particular, we report results in \gaia's synthetic HST/ACS F814W magnitudes, based on unweighted Sobel filters, and based on a method designed to avoid LF smoothing bias. To ensure the best match to extragalactic RG populations, we recommend using the \sargs-based TRGB calibrations reported here and in A24 to ensure the best available accuracy while also ensuring the best consistency with the target populations at greater distances aside from the caveats listed in the previous paragraph.

We caution that distance measurements based on our calibrations must follow an equivalent methodology in order to avoid methodological bias, notably with respect to sample selection. In particular, our \Bseq\ calibration can only be used to measure distances to galaxies where the distinction of PL-sequences is possible. Furthermore, accurate TRGB distances require the careful application of consistent algorithmic choices, such as smoothing scale and Sobel filter weighting, which can influence the results at the level of approximately $0.06$\,mag ($3\%$ in distance) as described in A24.

\section{Summary and conclusions \label{sec:conclusions}}

We have investigated the SMC's TRGB using variability-selected subsamples and showed that all RGs near the TRGB measured inside the OGLE-III SMC footprint exhibit small-amplitude long-period pulsations (Fig.\,\ref{fig:CMD}). Since the same is true in the LMC (A24), we conjecture that this type of variability is an astrophysical property of all high-luminosity red giants. Furthermore, we found that SMC RGs on sequence A are younger than those on sequence B (Fig.\,\ref{fig:Age}), just as in the LMC, and just as predicted by the evolutionary picture proposed by \citet{Wood15}. 

Comparing homogeneously determined ages of SARGs in the LMC and SMC based on APOGEE spectra \citep{Povick2023,Povick2023SMC}, we found that a) the SMC RGs are younger than LMC RGs, despite their lower metallicity and b) that the amplitude of pulsations in the SMC SARGs tends to be lower than in the LMC, especially for the B-sequence. The resulting age-amplitude relation (Fig.~\ref{fig:AgeAmplitude}) may offer a useful avenue for assessing the ages of RG populations.

The SMC's P-L sequences of long-period variables are shifted to significantly shorter periods compared to the LMC. This is readily explained by the reduced opacity of lower metallicity stars. Hence, SARG pulsation periods carry valuable information for standardizing TRGB magnitudes according to differences in chemical composition. As a consequence of this shift in period due to metallicity, RGs near the RGB tip obey a period-color relation (Fig.\,\ref{fig:PeriodcolorRelationship}), which we here report for the first time and at a statistical significance of $6-7\,\sigma$. These period-color relations could be useful for several other applications, e.g., for calibrating the TRGB method based on \gaia\ parallaxes.

The age and metallicity information derived from the variability of SARGs allows us to qualitatively trace the SMC's age gradient using the relative number of {\tt A-} to \Bseq\ SARGs. Additionally, the period-metallicity relation at the TRGB allowed us to trace a metallicity gradient in the SMC by the decrease in period for both A- and B-sequence RGs with increasing radius, matching the trend reported based on spectroscopy \citep{Povick2023SMC}. Using purely variability-derived information, we thus qualitatively showed that age increases with distance from the SMC's core, while metallicity decreases.

We measured the extinction-corrected TRGB apparent magnitude in the SMC for four subsamples and found that the \Bseq\ provides the most precise and robust (stable against methodological parameter choices) result. Additionally, we confirmed the hierarchy of \mtrgb\ values measured in the LMC (A24), with the \Bseq\ sample yielding the brightest \mtrgb\ of \mih$=14.920 \pm 0.010 (\mathrm{stat.}) \pm 0.008 (\mathrm{syst.})$\,mag using \gaia\ synthetic ACS/F814W photometry. Using OGLE-III photometry, we find \mio$=14.911 \pm 0.009 (\mathrm{stat.}) \pm 0.008 (\mathrm{stat.})$\,mag, and we show that the small difference between \mio\ and \mih\ is fully consistent with the differences in the photometric systems at the SMC's TRGB color (Fig.\,\ref{fig:GaiaVsOGLEIband}). 

Assuming the known DEB distance of the SMC \citep{Graczyk2020}, we obtained the most accurate TRGB calibration for the SMC's \Bseq, \Mih$= -4.057 \pm 0.019 (\mathrm{stat.}) \pm 0.029 (\mathrm{syst.})$\,mag (total error $1.5\%$ in distance) using synthetic {\it HST} ACS/F814W magnitudes derived from \gaia's synthetic photometry. The other stellar samples were found to be rather sensitive to the choice of smoothing parameters, resulting in significantly lower precision. Thus, the \Bseq\ results are best suited for measuring the difference in distance modulus between the Magellanic Clouds and for investigating the effects of differences in metallicity and reddening, the latter of which are readily excluded due to small color excess. 

We found that the color-based metallicity correction by \citet{Rizzi2007} brings the \Bseq-based absolute TRGB calibrations in both Clouds into near perfect agreement. However, the small difference in metallicity between LMC and SMC RGs limits the statistical significance of this result. Adopting the distance modulus difference from DEBs from \citet{Graczyk2020}, we determined a period-based metallicity correction that creates agreement between \Mio\ in the LMC and SMC, and we found that converting this period-based correction back to a color-based correction using the period-color relation yields a  result consistent with the literature. We consider this a useful first step and consistency check and note that a larger metallicity lever is needed to further calibrate a period-based metallicity correction for the TRGB. 

Our results highlight the systematic uncertainties due to RG population diversity inherent in TRGB distance measurements. In particular, we caution that our (metallicity-dependent) \Bseq\ calibration can only be used for distance measurements when detailed variability information is available to perform analogous sample selections. When this is not the case, we recommend using the \sargs\ based calibration, which is based on all stars near the tip of the RGB. Unfortunately, the precision of the SMC's \sargs-based calibration is negatively affected by the sensitivity of its LF to smoothing bias (\Mih$= -4.024 \pm 0.041 (\mathrm{stat.}) \pm 0.029 (\mathrm{syst.})$\,mag). A more accurate \sargs-based calibration based on the LMC was presented by A24.
We stress the need to apply consistent methodologies for TRGB calibration and distance measurements, notably viz. Sobel filter weighting and LF smoothing, to avoid bias on the order of $\sim 3\%$ (cf. A24).

Going forward, we believe that the variability of RGs near the TRGB will provide further useful insights into TRGB measurements and calibration. Further study is needed, notably involving spectroscopy and larger samples of SARGs. Even if the low-level variability of SARGs is difficult to measure in detail beyond the local group, understanding the impact of population diversity and population differences on extragalactic distances is crucial to obtaining unbiased TRGB-calibrated $H_0$ measurements.

\begin{acknowledgments}
We thank the anonymous referee for a constructive and detailed report that resulted in an improved manuscript. We thank Rick Hessman, Adam G. Riess, and Michele Trabucchi for useful discussions.
NWK acknowledges support from a ThinkSwiss Fellowship and the EPFL Excellence in Research Internship Program. RIA is funded by the Swiss National Science Foundation (SNSF) through an Eccellenza Professorial Fellowship (award PCEFP2\_194638). This project has received funding from the European Research Council (ERC) under the European Union's Horizon 2020 research and innovation programme (Grant Agreement No. 947660). 
\end{acknowledgments}

\facilities{OGLE, Gaia}

\bibliography{biblio}{}
\bibliographystyle{aasjournal}

\end{document}